\documentclass[%
 reprint,
 superscriptaddress,
 preprintnumbers,
 amsmath,amssymb,
 aps,
 showkeys,
 prd,
 onecolumn,
 notitlepage,
]{revtex4-1}

\usepackage{bm}
\usepackage{subfig}
\usepackage{xspace}

\usepackage{graphicx}                                  
\DeclareGraphicsExtensions{.eps,.ps,.eps.gz,.ps.gz}   
\graphicspath{{figs/}}

\def\goes{\rightarrow}

\newcommand{\zee}{$Z\rightarrow e^+e^-$}

\newcommand{\Et}{E_{\rm T}}

\newcommand{\Pt}{{\rm p}_{\rm T}}
\newcommand{\pt}{$\Pt$}
\newcommand{\vecpt}{\vec{\rm p}_{\rm T}}

\newcommand{\zmumu}{$Z\rightarrow\mu^+\mu^-$}

\newcommand{\ppbar}{$p\overline{p}$}

\newcommand{\lum}{4.6 $\mathrm{fb^{-1}}$}

\newcommand{\degs}{\mbox{$^{\circ}$}}

\newcommand{\plus}{\kern -0.18em +\kern -0.18em}
\newcommand{\MeV}{\ensuremath{\mathrm{\ Me\kern -0.1em V}}\xspace}
\newcommand{\MeVc}{\ensuremath{\mathrm{\ Me\kern -0.1em V\kern -0.1em  \mathit{c}}}\xspace}
\newcommand{\MeVcsq}{\ensuremath{\mathrm{\ Me\kern -0.1em V\kern -0.1em \mathit{c}^2}}\xspace}
\newcommand{\GeV}{\ensuremath{\mathrm{Ge\kern -0.1em V}}\xspace}
\newcommand{\GeVc}{\ensuremath{\mathrm{\ Ge\kern -0.1em V\kern -0.1em \mathit{/c}}}\xspace}
\newcommand{\GeVcsq}{\ensuremath{\mathrm{\ Ge\kern -0.1em V\kern -0.1em \mathit{/c}^2}}\xspace}
\newcommand{\TeV}{\ensuremath{\mathrm{Te\kern -0.1em V}}\xspace}
\newcommand{\tev}{\ensuremath{\mathrm{Te\kern -0.1em V}}\xspace}
\newcommand{\bfTeV}{\ensuremath{\bf{Te\kern -0.1em V}}\xspace}
\newcommand{\nsGeV}{\ensuremath{\mathrm{Ge\kern -0.1em V}}\xspace}
\newcommand{\nsGeVc}{\ensuremath{\mathrm{Ge\kern -0.1em V\kern -0.1em \mathit{c}}}\xspace}

\begin{document}

\preprint{FERMILAB-PUB-10-280-PPD}
\preprint{EFI-10-17}
\preprint{ANL-HEP-PR-10-41}

\title{Present Limits on the Precision of SM Predictions for Jet Energies}

\affiliation{Argonne National Laboratory, Argonne, IL 60439, USA}
\affiliation{Enrico Fermi Institute, University of Chicago, Chicago, IL 60637, USA}
\affiliation{University of Liverpool, Liverpool L69 7ZE, UK}
\affiliation{Fermi National Accelerator Laboratory, Batavia, IL 60510, USA}
\author{A.A.~Paramonov} 
\affiliation{Argonne National Laboratory, Argonne, IL 60439, USA}
\author{F.~Canelli}
\affiliation{Enrico Fermi Institute, University of Chicago, Chicago, IL 60637, USA}
\author{M.~D'Onofrio}
\affiliation{University of Liverpool, Liverpool L69 7ZE, UK}
\author{H.J.~Frisch} 
\affiliation{Enrico Fermi Institute, University of Chicago, Chicago, IL 60637, USA}
\author{S.~Mrenna} 
\affiliation{Fermi National Accelerator Laboratory, Batavia, IL 60510, USA}

\begin{abstract}
We investigate the impact of theoretical uncertainties on the accuracy
of measurements involving hadronic jets. The analysis is performed
using events with a $Z$ boson and a single jet observed in $p\bar{p}$
collisions at $\sqrt{s}$ = 1.96 TeV in \lum~of data from the Collider
Detector at Fermilab (CDF). The transverse momenta (\pt) of the jet
and the boson should balance each other due to momentum conservation
in the plane transverse to the direction of the $p$ and $\bar{p}$
beams. We evaluate the dependence of the measured \pt-balance on
theoretical uncertainties associated with initial and final state
radiation, choice of renormalization and factorization scales, parton
distribution functions, jet-parton matching, calculations of matrix
elements, and parton showering. We find that the uncertainty caused by
parton showering at large angles is the largest amongst the listed
uncertainties. The proposed method can be re-applied at the LHC
experiments to investigate and evaluate the uncertainties on the
predicted jet energies. The distributions produced at the CDF
environment are intended for comparison to those from modern event
generators and new tunes of parton showering.
\end{abstract}

\keywords{Jet Energy Scale; JES; Collider Detector at Fermilab; CDF;
Fermilab; Parton Shower; Final State Radiation; FSR}

\maketitle

\section{\label{intro}Introduction and physics motivation}
%
\par The discovery potential of LHC experiments will strongly depend
on the accuracy of standard model (SM) predictions for processes
containing hadronic jets~\cite{Meirose}, as the first step in
establishing ``new physics'' has to be identifying an incontrovertible
deviation from SM phenomena.  The uncertainties on predictions for SM
processes directly impact measurements of jet spectra, searches for
new heavy particles using jet energies for kinematic reconstruction,
and the calculation of missing transverse momentum, to name several
prevalent analysis strategies as examples. The discovery potential for
supersymmetry and other models of physics beyond the SM thus relies on
having calibrated methods for measuring jet energies.

\par It has been common practice to normalize the clustered jet
energy, measured with calorimeters, to the energy of the particle jet
or the parent parton~\cite{jet_corr,D0_JES}. The correction factor is
often called the jet energy scale (JES). At CDF jets are observed in
non-compensating sampling calorimeters, which have a non-linear
response to single particles. The simulated calorimeter response for
single hadrons is tuned to match that in data~\cite{jet_corr}. The
measured jet energy is corrected for instrumental effects such as the
non-linear response of the calorimeters as well as for parton
radiation and hadronization effects. The correction for radiation and
hadronization effects is independent of the experimental setup.  In
addition, the jet energies are corrected for multiple
\ppbar~interactions in the same bunch crossing.

\par The systematic uncertainties on the JES and the related
measurements arise from the accuracy of the detector simulation and
limitations of the methods used by SM event generators. The event
generators, such as {\sc pythia} and {\sc
alpgen}~\cite{Pythia,Alpgen}, use a simplified modeling of complex SM
processes that can be altered by tuning internal parameters. The
model-dependent aspects we investigate are the following:
\begin{itemize}
\item parton distribution functions (PDFs) of the colliding $p$ and
  $\bar{p}$
\item leading order (LO) matrix elements of tree-level processes
  such as $q\bar{q}\goes Zg$ and $qg\goes Zq$
\item the parton-jet matching scheme~\cite{MLM_matching}
\item final state radiation (FSR)
\item initial state radiation (ISR)
\item the renormalization and factorization scales
\item residual effects due to multiple \ppbar~interactions
\item the ability of the leading-log parton showering model to
describe radiation at large angles.
\end{itemize}

\par To perform the analysis we select events with a $Z$ boson and a
jet observed in \lum~of data from CDF. A $Z$ boson is clearly
identified as a pair of opposite-sign electrons or muons with an
invariant mass close to the $Z$-boson mass. The transverse momentum of
the boson is measured with high precision so that the $Z$+jet sample
is ideal for the analysis. We use the \pt-balance in the event, with
the mean-value of the ratio \pt$(jet)$/\pt$(Z)$ as the observable of
interest, to test the simulated SM predictions.

\par The determination of the jet energy scale used in
previously-published CDF analyses was performed with about 300
pb$^{-1}$ of data~\cite{jet_corr}. The overall uncertainty on the JES
was compared to the difference in \pt-balance between data and MC
predictions for a photon-jet and a $Z$-jet samples; the difference in
\pt-balance was calculated in a similar fashion to the method used in
the current analysis. Having significantly more data (\lum) we
investigate the systematic uncertainties affecting measurements of jet
energies independently from the previously evaluated uncertainties on
the JES. Using the additional statistical power, we can disentangle
the different effects contributing to the uncertainties by correlating
the limitations in theoretical predictions with their effect on the
\pt-balance.

\par The outline of the paper is as follows. The CDF II detector is
described in Section~\ref{cdf_det}. Details for MC event generators
are presented in Section~\ref{montecarlo}. Event selection follows in
Sections~\ref{data} and~\ref{evt_sel}. We show \pt-balance in data and
{\sc pythia} as a function of \pt$(Z)$ in Section~\ref{QnG_jets}. In
the same section we further validate LO predictions from {\sc pythia}
and {\sc alpgen} by checking that the relative contributions from
$qg\goes Zq$ and $q\bar{q}\goes Zg$ diagrams are accurately modeled.
Computation-related uncertainties such those due to the choice of
factorization and renormalization scales are evaluated in
Sections~\ref{Py_Alp}, \ref{PDFs}, \ref{ISR}, \ref{Q2_Alp},
and~\ref{FSR}. We use data to evaluate uncertainties due to
mis-modeling of parton radiation at large angles and multiple
\ppbar~interactions in Sections~\ref{OOC_rad} and~\ref{pileup},
respectively. The uncertainty due to detector simulations is
calculated in Section~\ref{det_SPR}. In Section~\ref{sum_of_syst} we
summarize the observed uncertainties and compare those to the
difference in \pt-balance between data and the MC predictions. We
present conclusions in Section~\ref{conclusions}.

\section{\label{cdf_det}The CDF II detector}
\par The CDF II detector is a cylindrically symmetric spectrometer
designed to study \ppbar~collisions at the Fermilab Tevatron. The
detector has been extensively described in the
literature~\cite{CDFdetector}. Here we briefly describe the detector
subsystems relevant for the analysis.

\par Tracking systems are used to measure the momenta of charged
particles, and to trigger on and identify leptons with large
transverse momentum, $\Pt$~\cite{EtPt}. A multi-layer system of
silicon strip detectors (SVX)~\cite{SVX}, which identifies tracks in
both the $r-\phi$ and $r-z$ views~\cite{CDF_coo}, and the central
outer tracker (COT)~\cite{COT} are contained in a superconducting
solenoid that generates a magnetic field of 1.4 T. The COT is a 3.1 m
long open-cell drift chamber that makes up to 96 measurements along
the track of each charged particle in the region $|\eta|<1$. Sense
wires are arranged in 8 alternating axial and stereo ($\pm 2\degs$)
super-layers with 12 wires each. For high momentum tracks, the COT
$\Pt$ resolution is $\sigma_{\Pt}/\Pt^2 \simeq
0.0017$~$(\GeVc)^{-1}$~\cite{CDF_Momentum}.

\par Segmented calorimeters with towers arranged in a projective
geometry, each tower consisting of an electromagnetic and a hadronic
compartment~\cite{cem_resolution, cal_upgrade}, cover the central
region, $|\eta|<1$ (CEM/CHA), and the forward
region~\cite{Apollinari1998515}, $1<|\eta|<3.6$ (PEM/PHA). In both the
central and forward regions, systems with finer spatial resolution are
used to make profile measurements of electromagnetic showers at shower
maximum~\cite{CDFII} for electron identification (the CES and PES
systems, respectively).  Electrons are reconstructed in the CEM with
an $\Et$~\cite{EtPt} resolution of $\sigma(\Et)/\Et \simeq
13.5\%/\sqrt{\Et/\GeV}\oplus 2\%$~\cite{cem_resolution} and in the PEM
with an $\Et$ resolution of $\sigma(\Et)/\Et \simeq
16.0\%/\sqrt{\Et/\GeV}\oplus 1\%$~\cite{pem_resolution}. Jets are
identified using a cone clustering algorithm in $\eta-\phi$ space,
with cone radius $R=\sqrt{\Delta\eta^2+\Delta\phi^2}$, as a group of
electromagnetic and hadronic calorimeter towers. The CDF hadronic
calorimeters have a steel-scintillator sampling design and the
electromagnetic calorimeters are built from lead and scintillator. The
sampling calorimeters have a non-linear response to stable
hadrons~\cite{jet_corr}, which carry most of the jet momentum.

\par Muons are identified using the central CMU, CMP, and
CMX~\cite{cmu_ref, cmp_ref} muon systems, which cover the kinematic
region $|\eta|<1$. The CMU system uses four layers of planar drift
chambers to detect muons with $\Pt>1.4~\GeVc$ in the central region of
$|\eta|<0.6$. The CMP system consists of an additional four layers of
planar drift chambers located behind 0.6 m of steel outside the
magnetic return yoke, and detects muons with $\Pt>2.0~\GeVc$. The CMX
detects muons in the region $0.6<|\eta|<1.0$ with four to eight layers
of drift chambers, depending on the polar angle.

\par The beam luminosity is measured using two sets of gas Cherenkov
counters, located in the region $3.7<|\eta|<4.7$. The total
uncertainty on the luminosity is estimated to be 5.9\%, where 4.4\%
comes from the acceptance and operation of the luminosity monitor and
4.0\% from the calculation of the inelastic
\ppbar~cross-section~\cite{luminosity}.

\par A 3-level trigger system~\cite{CDFdetector} selects events for
further analysis offline.  The first two levels of triggers consist of
dedicated fast digital electronics analyzing a subset of the full
detector data.  The third level, applied to the full data from the
detector for those events passing the first two levels, consists of a
farm of computers that reconstruct the data and apply selection
criteria for (typically) several hundred distinct trigger paths.

\section{\label{montecarlo}Standard model predictions for events with
 a $Z$ boson and jets}

\par The standard model expectations for inclusive production of $Z$
bosons are calculated from Monte Carlo simulations using {\sc pythia}
and {\sc alpgen}. Events from the two MC generators are processed
through the full detector simulation to be reconstructed and analyzed
like data.
\par The datasets for the $Z$ + light jets signatures are produced
using v6.216 of {\sc pythia} in which the $\Pt$ spectrum of the $Z$
bosons, $\Pt^Z$, has been tuned to CDF Run I data for $0 < \Pt^Z < 20$
\GeVc~\cite{Willis_tuned_Pythia}, and which incorporates a tuned
underlying-event, Tune AW~\cite{Rick_Field}. The event generator was
set to inclusive production of $Z$-bosons with a $M(\gamma^*/Z)$ $>$
30 \GeVcsq cut. Historically, a di-jet sample simulated with {\sc
pythia} was used to determine the JES at CDF; in this study we take
the $Z$ + jets events from {\sc pythia} as our default benchmark
sample.
\par Additional $Z$ + jets samples are produced with v2.10-prime of
{\sc alpgen} that has built-in matching of the number of jets from
parton showering and matrix-element production~\cite{MLM_matching}.
The exclusive $Z$ + N partons (N=0,..,4) samples were combined into
one inclusive sample using the corresponding cross-sections provided
by {\sc alpgen}.  Showering and hadronization of jets is done with
{\sc pythia} v6.326, Tune AW~\cite{Rick_Field}. The jet-parton
matching is performed at a \pt~of 15 \GeVc (referred to as the
matching scale) using the {\sc jetclu} clustering algorithm with a
radius of R=0.4.
\par Production of $Z$+jet events is performed differently by
stand-alone {\sc pythia} and {\sc alpgen+pythia}. The {\sc
alpgen+pythia} calculation begins with the exact matrix elements from
{\sc alpgen} for $Z$+N partons (N=0,..,4), which are then interfaced
with {\sc pythia} parton showering. The interface contains a veto
algorithm that removes double-counting between matrix element and
parton shower partons~\cite{MLM_matching}. The stand-alone {\sc
pythia} calculation begins with the simplest matrix element ($Z$+0
partons) and adds additional partons from the shower with no need for
a veto. However, the first parton emission is corrected to reproduce
the $Z$+1 parton matrix element. Thus, any substantial differences
between the predictions of the two calculations (if they exist) should
arise for the second jet. Both event generators use the same PDF set,
CTEQ5L, and Lund string hadronization model as implemented in Pythia.

\par The parton showering (PS) inside the jet cone (see
Section~\ref{jet_id}) has been extensively
studied~\cite{QCD_part_kT,QCD_charged_pt,PhysRevD.71.112002} and is in
good agreement with predictions. In addition, the momentum spectra of
charged-particle tracks in jets are found to be in good agreement with
SM predictions~\cite{jet_corr}.

\par The large-angle (outside of cone R=0.4) parton radiation is not
described well by stand-alone parton showering model in {\sc pythia};
the rate of softer jets collinear to a jet is not described (e.g. see
Fig. 3 in~\cite{Bruce_VISTA}, the distribution of $\Delta R(j2,j3)$
observed in multi-jet events). Radiation of the 3rd jet in multi-jet
events is qualitatively equivalent to radiation of the 2nd jet in
$Z$+jets events. The same issue affects the invariant mass calculated
for a pair of jets, energy of a jet, and missing transverse momentum.
The problem related to the parton radiation at large angles can be
addressed by using an exact matrix element (ME) for multi-jet events
as is done in {\sc alpgen+pythia} simulation. The ME correction is
introduced for radiated jets with \pt~above the matching scale
(15~\GeVc); softer radiation is produced via the same leading-log
parton shower mechanism.

\section{\label{data}Description of data samples}
\par The analysis uses events that contain either an electron with 
$\Et>18$ \GeV or a muon with $\Pt>18$ \GeVc selected within the
central region of the detector, $|\eta|<1$, by the trigger system. The
electron dataset contains 229M events; the muon dataset contains about
65M events. The integrated luminosity of each dataset is~\lum.

\section{\label{evt_sel}Event selection}
\par Both the observed and the simulated events (see
Section~\ref{montecarlo}) use the same selection criteria to identify
electrons, muons, $Z$ bosons, and jets. Details of the selection
criteria for electrons and muons are provided in~\ref{lepton_id}.

\subsection{\label{jet_id}Jet identification}
\par Jets are reconstructed using {\sc jetclu}, the standard CDF
cone-based clustering algorithm, with cone radii of R = 0.4, 0.7, and
1.0~\cite{jet_clu}. The clustering is performed using calorimeter
towers with raw (uncorrected) energy above 1 \GeV to form a cluster of
at least 3 \GeV. To resolve ambiguities with overlapping cones, cones
sharing an energy fraction of more than 0.75 are merged into a single
jet; otherwise the shared towers are assigned to the closest jet.
\par The jet energies are corrected for the non-uniformity in $\eta$
of the calorimeter response and for multiple \ppbar~interactions. In
this analysis, the leading jet energy is always corrected to the
parton level; the jet energy scale is adjusted to relate the measured
energy of a simulated jet and the energy of the corresponding parton
in di-jet events~\cite{jet_corr} from {\sc pythia}. The jet clustering
algorithm is run over calorimeter towers for reconstructed jets, and
over stable particles for hadron-level jets. The correction factor
from the hadron to the parton level is a function of only jet \pt, and
is the same for data and predictions. The JES corrections at CDF do
not take into account if a jet is initiated by a tree-level quark or a
gluon.


\par Calorimeter clusters that coincide with an identified electron,
or photon are removed from the jet collection to avoid ambiguities.
High-$\Pt$ photons are not rare in hard-scattering events. Identifying
photons as jets and then correcting them as jets can lead to
mis-measured \pt-balance. To avoid photon misidentifications the event
selection requires the leading jet to have EM-fraction less than
0.95. The EM-fraction is the fraction of energy of a jet deposited in
the electromagnetic compartment of the calorimeter in comparison to
the total energy of the jet.

\subsection{Reconstruction of $Z$ + jet events}
\par
\begin{figure}[h]
\centering
\subfloat[][]{\includegraphics[angle=0,width=0.45\textwidth]{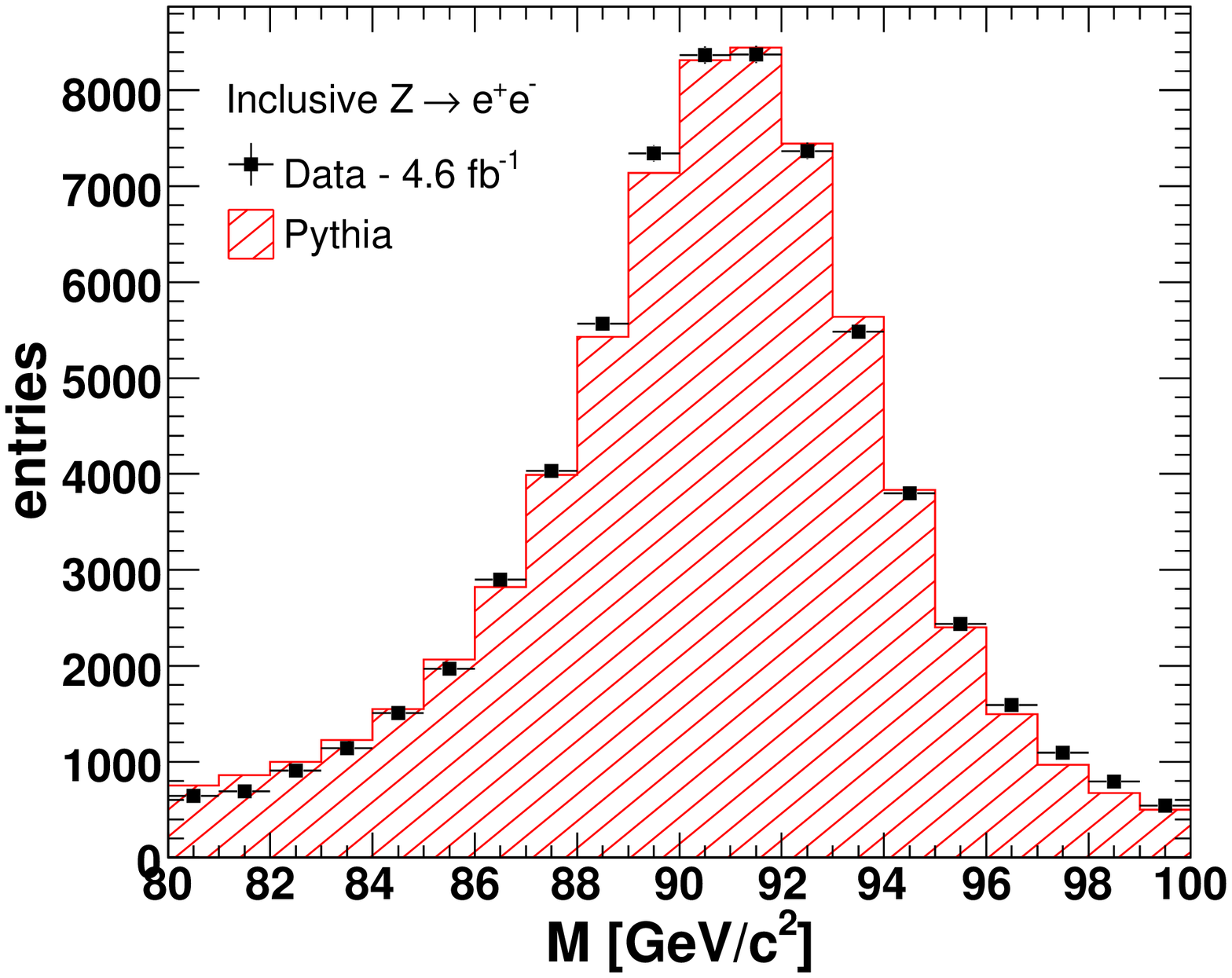}}
\subfloat[][]{\includegraphics[angle=0,width=0.45\textwidth]{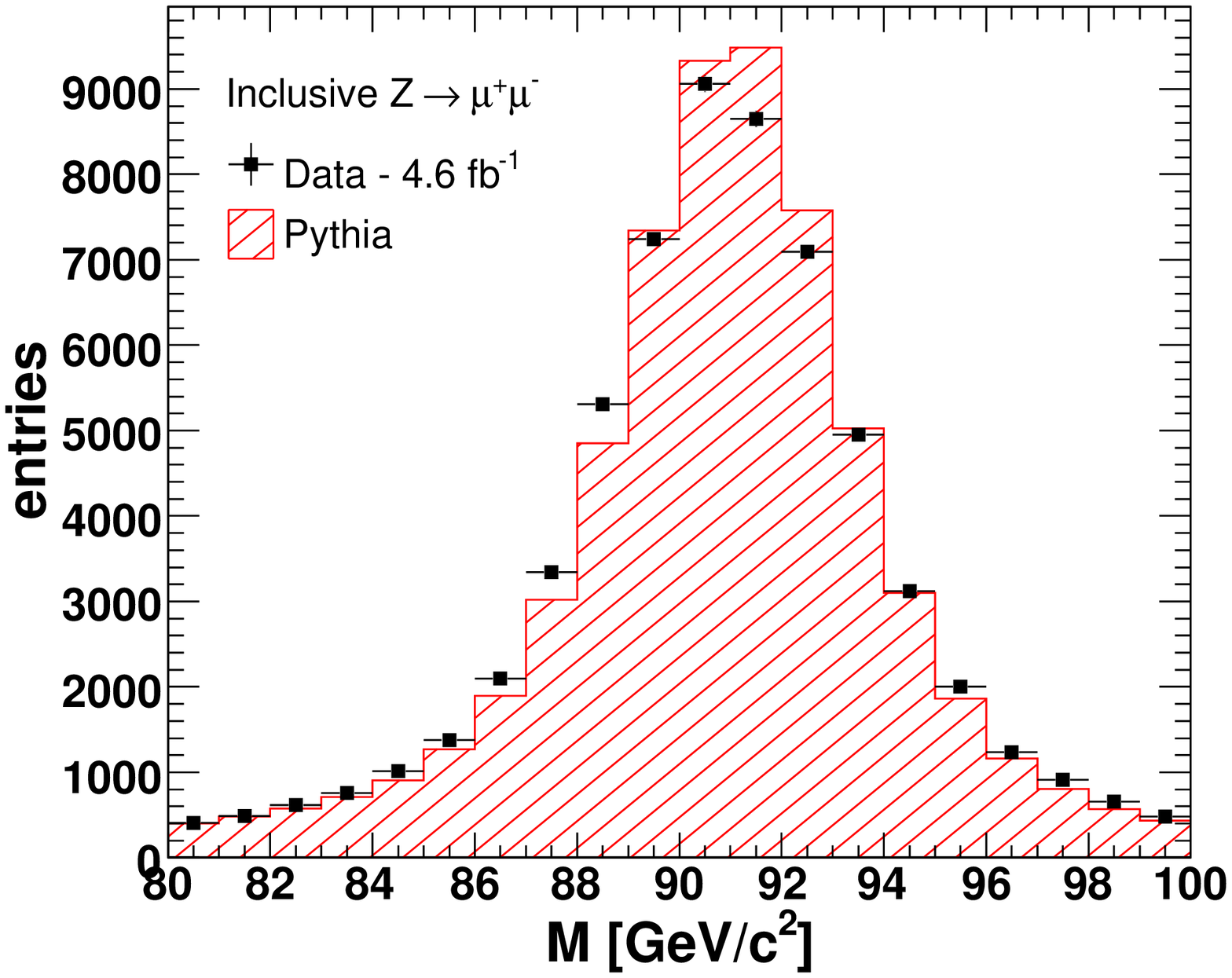}}
\caption{The observed (points) and expected (histogram) distributions
in the invariant mass of $e^+e^-$ (a) and $\mu^+\mu^-$ (b) lepton
pairs. The slight mismatch in data and Monte Carlo in the dimuon
spectrum is due to a small miscalibration in the standard CDF tracking
momentum scale.}
\label{fig:z_m}
\end{figure}

 Pairs of oppositely-charged electrons and muons are identified as
$Z$-boson candidates if the reconstructed invariant mass falls in the
mass window from 80~\GeVcsq to 100~\GeVcsq. The selection of
$Z\goes\ell\ell$ events requires two tight leptons or a tight and a
loose lepton (see~\ref{lepton_id}). The two leptons are required to be
assigned to the same primary vertex, which is required to have a
z-coordinate within 60 cm from the center of the CDF detector. Also we
remove dimuon events where the leading jet overlaps ($\Delta~R<0.4$)
with one of the muons forming a $Z$ boson.  Figure~\ref{fig:z_m} shows
the distributions in invariant mass for electron and muon pairs; the
data are in a remarkable agreement with SM predictions from {\sc
pythia} and {\sc alpgen}.
\par Events are further required to have at least one jet. First, we
correct all jet energies for $\eta$-dependent response of the
calorimeters and for multiple \ppbar~interactions; the leading jet
\pt~is required to be greater than 8~\GeVc. An event is vetoed if the
second jet cluster, the sub-leading jet, has a \pt~of more than
8~\GeVc. The leading jet is required to be in the pseudo-rapidity
range of $0.2<|\eta_{\rm det.}|<0.8$ to avoid cracks in the central
calorimeter. We do not apply the $\eta$ requirement to sub-leading
jets; their pseudo-rapidity can be from -2.8 to 2.8. Then the momentum
of the leading jet is corrected to the parton level as described above
(see Sec.~\ref{jet_id}).  The $\vecpt$ of the leading jet,
$\vecpt(jet1)$, and the $\vecpt$ of the $Z$ boson, $\vecpt(Z)$, are
required to be back-to-back: $\Delta\phi(\vecpt(jet1),\vecpt(Z))>3.0$
rad.

\par Rarely the leading jet can originate from another
\ppbar~interaction produced in the same bunch crossing as the $Z$+jet;
the overlapping jets bias the \pt-balance. The number of interactions
in each event is estimated via the number of primary vertices along
the beam line; the number is used for the corresponding JES
correction. The multiple interactions are mostly minimum bias events
that produce relatively soft jets (\pt~$\lesssim$ 8 \GeVc).

\par We veto events with two or more primary vertices in which the
leading jet is measured to come from a different interaction vertex
than that of the hard interaction producing the $Z$-boson. The jet
vertex of origin is determined using tracks pointing to the towers in
the jet cluster. For each track we take the $z$-coordinate of the
point on the track closest to the beam-line. A mean value of the
$z$-coordinates is calculated to determine the vertex of jet origin.
Specifically, events are removed if the leading jet has two or more
tracks, and the jet vertex is more than 2 cm away from the vertex of
the lepton pair along the $z$-axis. The veto has a negligible effect
on the \pt-balance as the leading jet \pt~is required to be above
8~\GeVc.

\par
\begin{figure}[h]
\centering
\includegraphics[angle=0,width=0.45\textwidth]{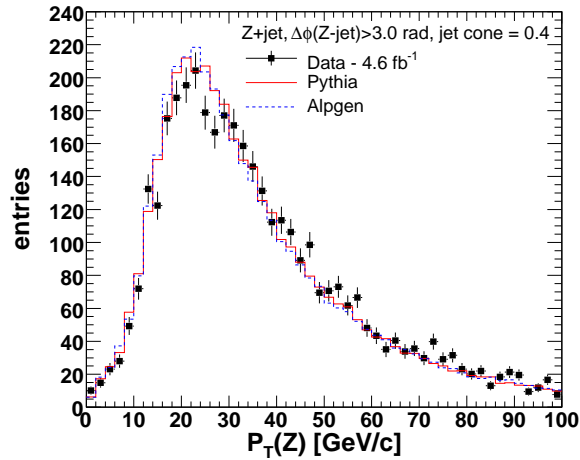}
\caption{The observed (points) and expected (histogram) distributions
in the transverse momentum of lepton pairs with invariant mass between
80$<M(\ell\ell)<$100~\GeVcsq. The solid line is for the {\sc pythia}
and the dashed line is for the {\sc alpgen} predictions. All other
sample selection cuts have been applied.}
\label{fig:z_pt}
\end{figure}

 We use simulations to evaluate the bias between the measured 
\pt$(Z)$ and its true  value separately for \zee~and \zmumu~events. 
The mean value of the observed \pt$(Z)$ is found to be within 0.5\% of
its mean generated value in both samples. The distribution of the
observed transverse momentum of $Z$-bosons, \pt$(Z)$, is shown in
Fig.~\ref{fig:z_pt}.

\par
\begin{figure}[h]
\centering
\subfloat[][]{\includegraphics[angle=0,width=0.45\textwidth]{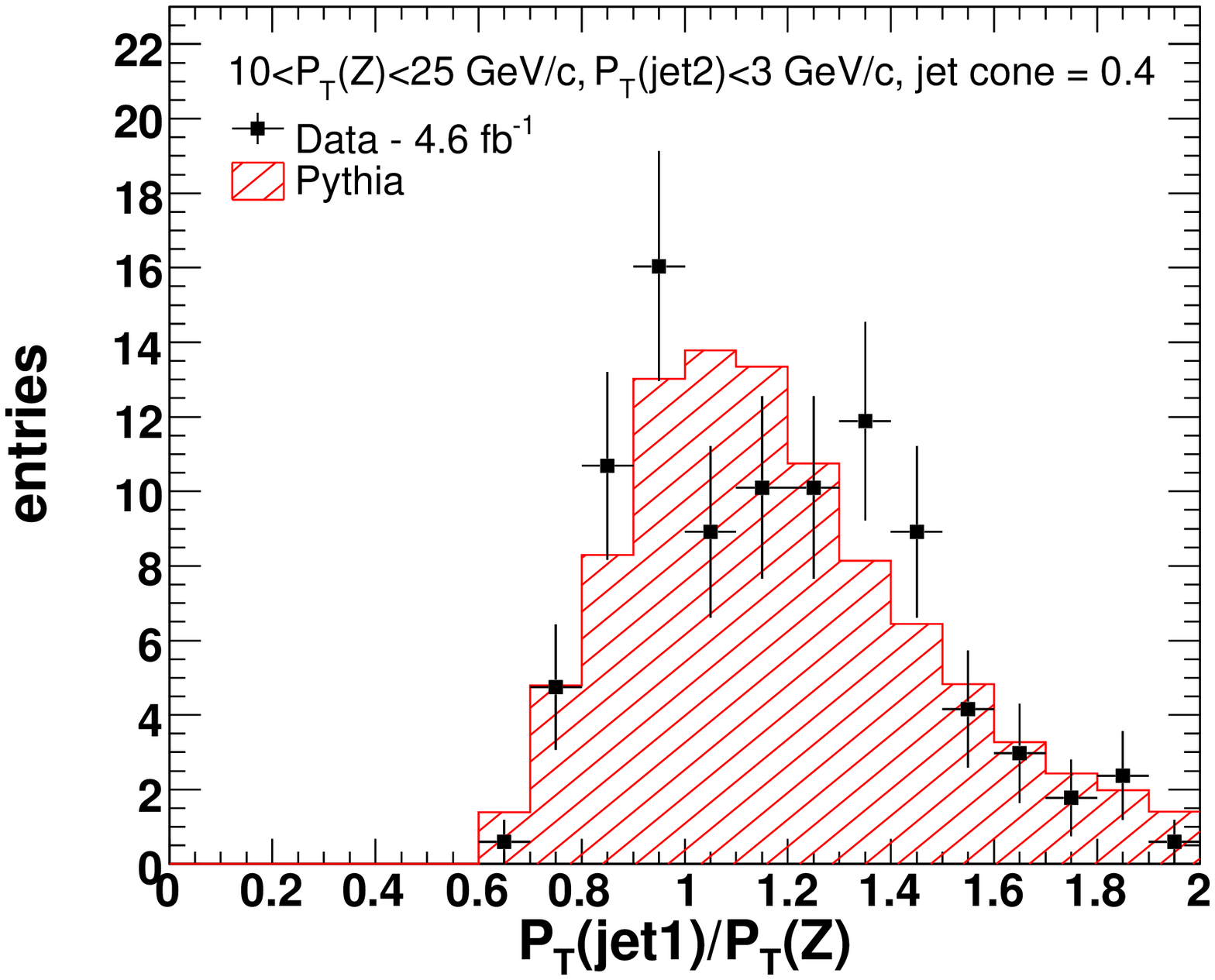}}
\subfloat[][]{\includegraphics[angle=0,width=0.45\textwidth]{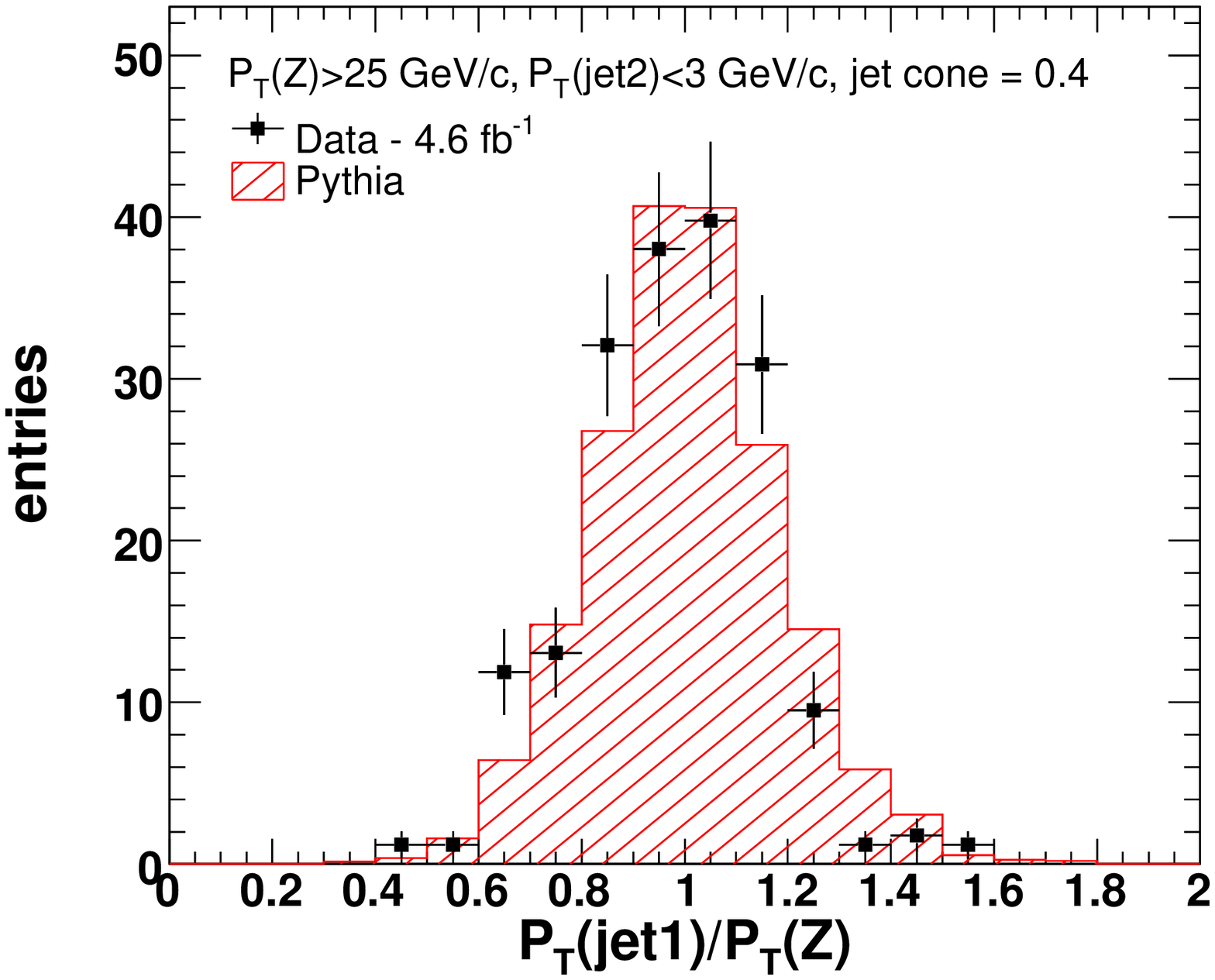}}
\caption{The observed (points) and expected (histogram) distributions
in \pt-balance, \pt$(jet1)$/\pt$(Z)$, for events with \pt$(Z)$ less
than 25~\GeVc (a) and greater than 25~\GeVc (b). Sub-leading jets are
required to have \pt~$<$ 3~\GeVc to suppress energy flow outside of
the cone of the leading jet (see Section~\ref{jet_id}). The
distribution (a), for \pt$(Z)$ $<$ 25~\GeVc, illustrates the turn-on
curve for the 8 \GeVc threshold on the \pt~of the leading jet,
\pt$(jet1)$ $>$ 8 \GeVc. }
\label{fig:balance}
\end{figure}

 The $Z$-jet system is not a perfect two-body process and the
\pt-balance, \pt$(jet1)$/\pt$(Z)$, is sensitive to the surroundings of
the jet. Also, the jet energy resolution is rather poor (in comparison
to the jet energy) for jets with \pt~of about 10~\GeVc. The resolution
improves with higher jet energies from approximately 20\% for \pt(jet)
= 40 \GeVc to $\sim$12\% for \pt(jet) = 140 \GeVc. As an illustration
that we can approach the ideal two-body system we apply a more
stringent event selection; the \pt~of sub-leading jets is required to
be less than 3~\GeVc. The \pt-balance for the exclusive event
selection is shown in Fig.~\ref{fig:balance}. The distribution for
events with \pt$(Z)$ $<$ 25~\GeVc is asymmetric and shifted from 1.0
due to the finite jet energy resolution and the cut-off on the minimum
\pt~of the leading jet (see Fig.~\ref{fig:balance}~(a)). The
distribution in \pt-balance for events with \pt$(Z)$ $>$ 25~\GeVc is
nearly symmetric and peaks close to 1.0 (see
Fig.~\ref{fig:balance}~(b)). Consequently we use events with \pt$(Z)$
$>$ 25~\GeVc to compare data and predictions. In the following
analysis, however, we do not use the 3~\GeVc cut-off on the
sub-leading jet \pt~as it was used to effectively prove the jet energy
correction works properly, but relax the cut-off to 8~\GeVc to better
study the systematic uncertainties caused by the MC predictions.

\section{\label{QnG_jets}Validation of SM simulations: Properties of quark and gluon jets}

\par Properties of a QCD jet depend on the tree-level parton
initiating it. A jet initiated by a gluon has a higher multiplicity of
daughter hadrons than a jet of the same energy initiated by a light
quark. The difference in the observed particle multiplicities is due
to the different color charges of quarks and
gluons~\cite{Derrick1985449}.
\par
\begin{figure}[h]
\centering
\subfloat[][]{\includegraphics[angle=0,width=0.49\textwidth]{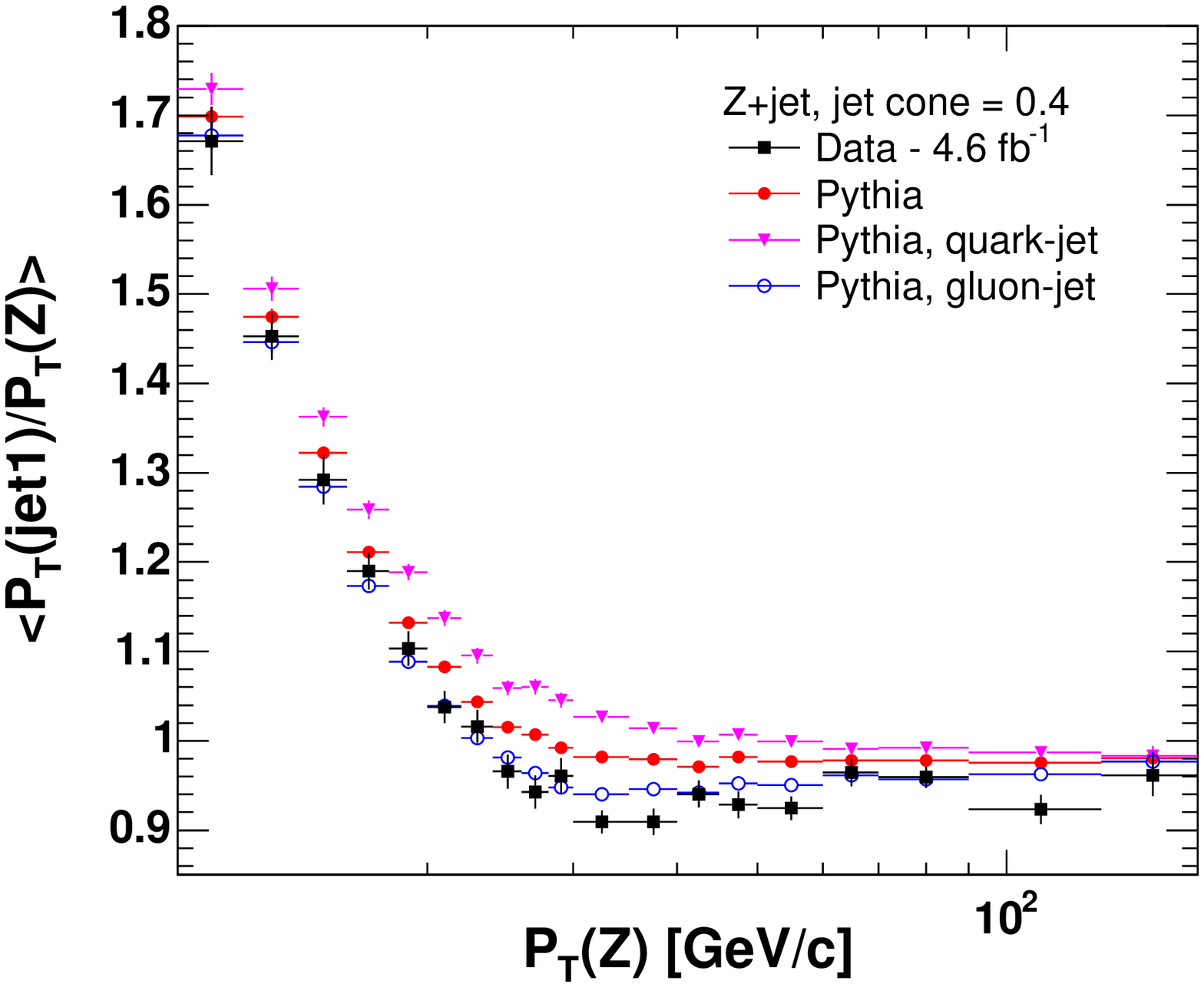}}
\subfloat[][]{\includegraphics[angle=0,width=0.49\textwidth]{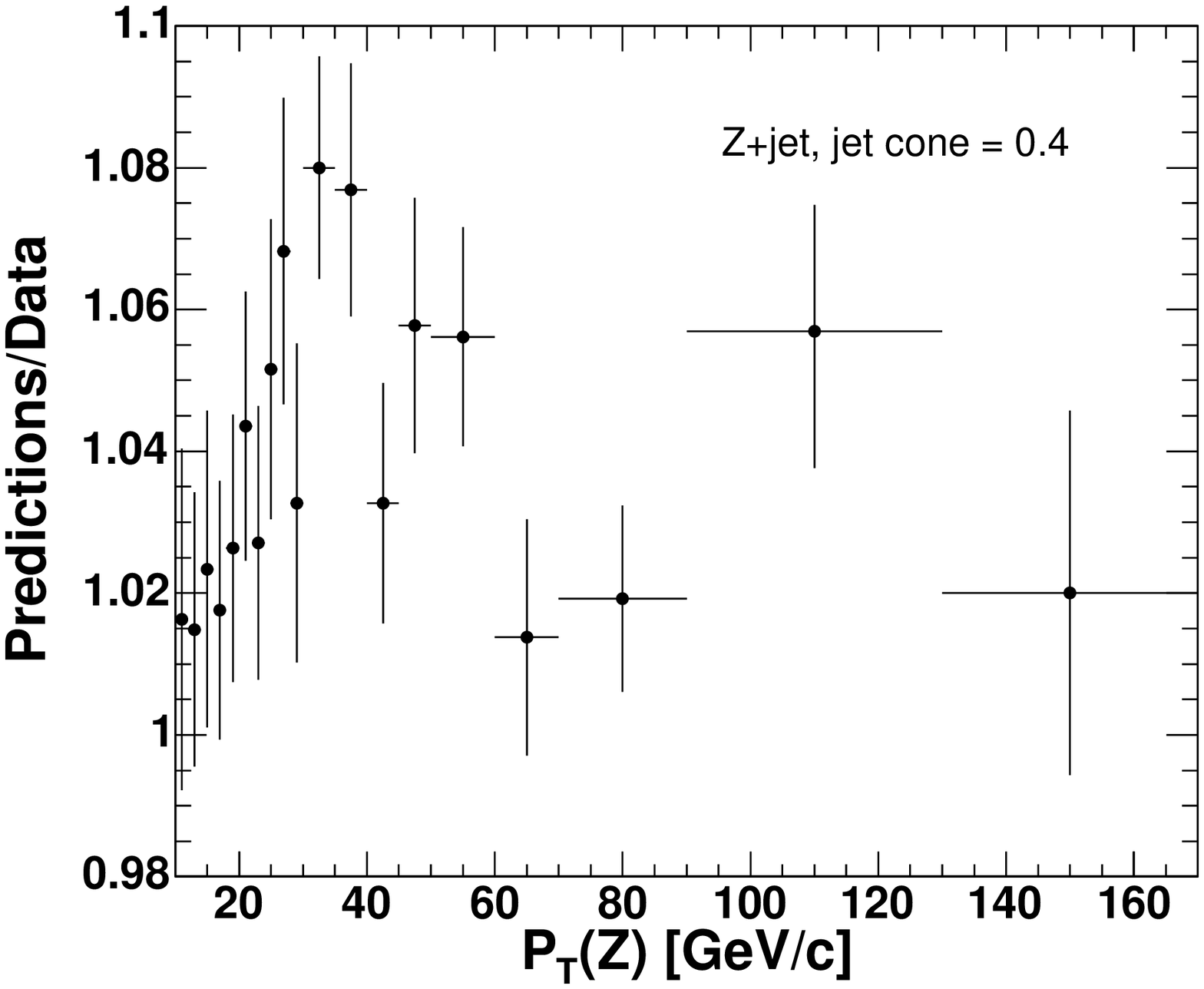}}
\caption{ a.) The average \pt-balance as a function of \pt($Z$). b.)
The ratio of predicted and measured distributions in \pt-balance. The
predicted distribution is for the combination of quark and gluon jets
given by {\sc pythia}. The jets are clustered using a cone radius of
R=0.4. The \pt-balance is noticeably distorted by the cut-off on the
minimum \pt~of the leading jet and a finite jet energy resolution in
events with \pt$(Z)$ $<$ 25 \GeVc. These events will not be used for
the study of the predicted jet energy.}
\label{fig:balance_cone04}
\end{figure}

\begin{figure}[h]
\centering
\subfloat[][]{\includegraphics[angle=0,width=0.49\textwidth]{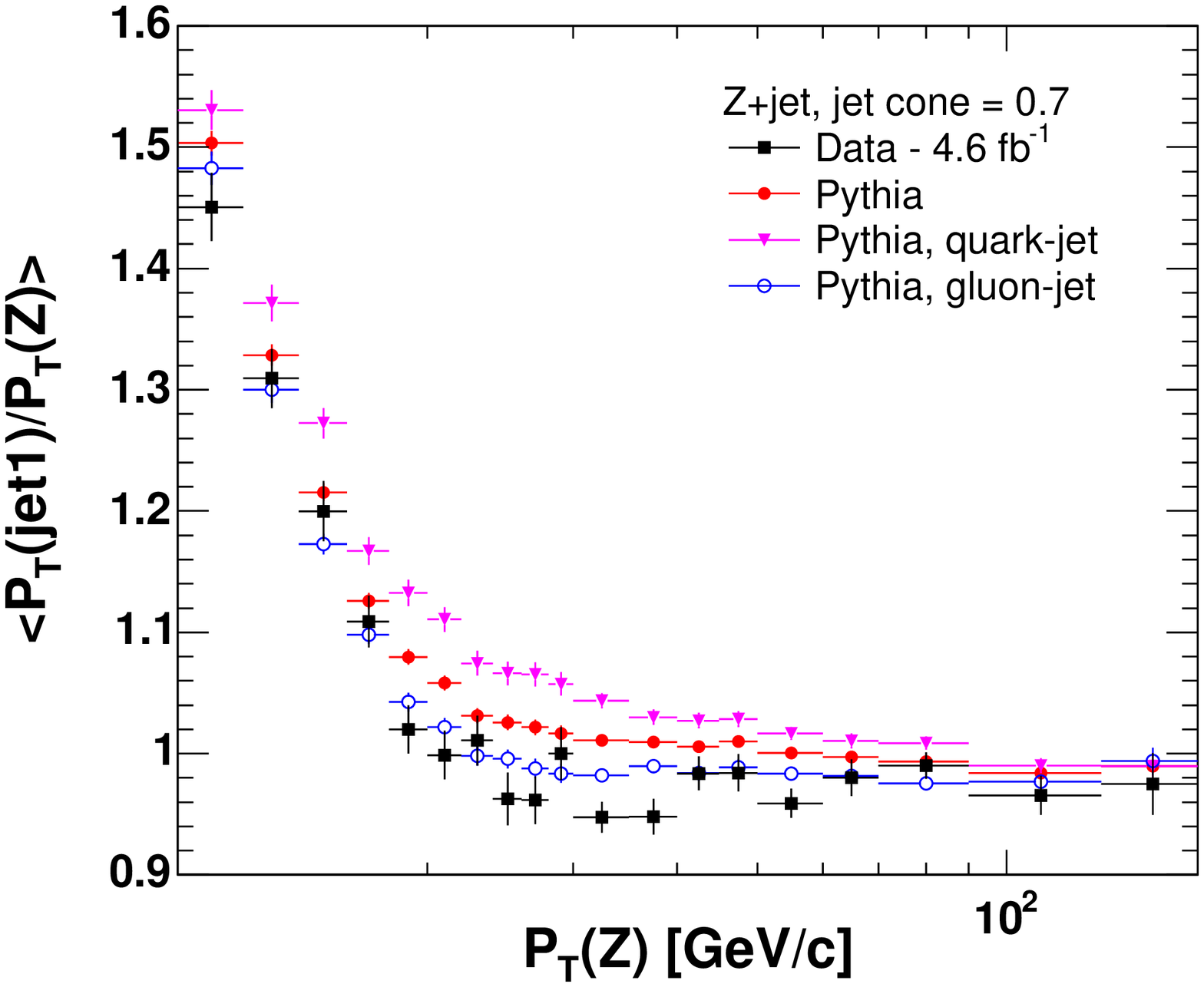}}
\subfloat[][]{\includegraphics[angle=0,width=0.49\textwidth]{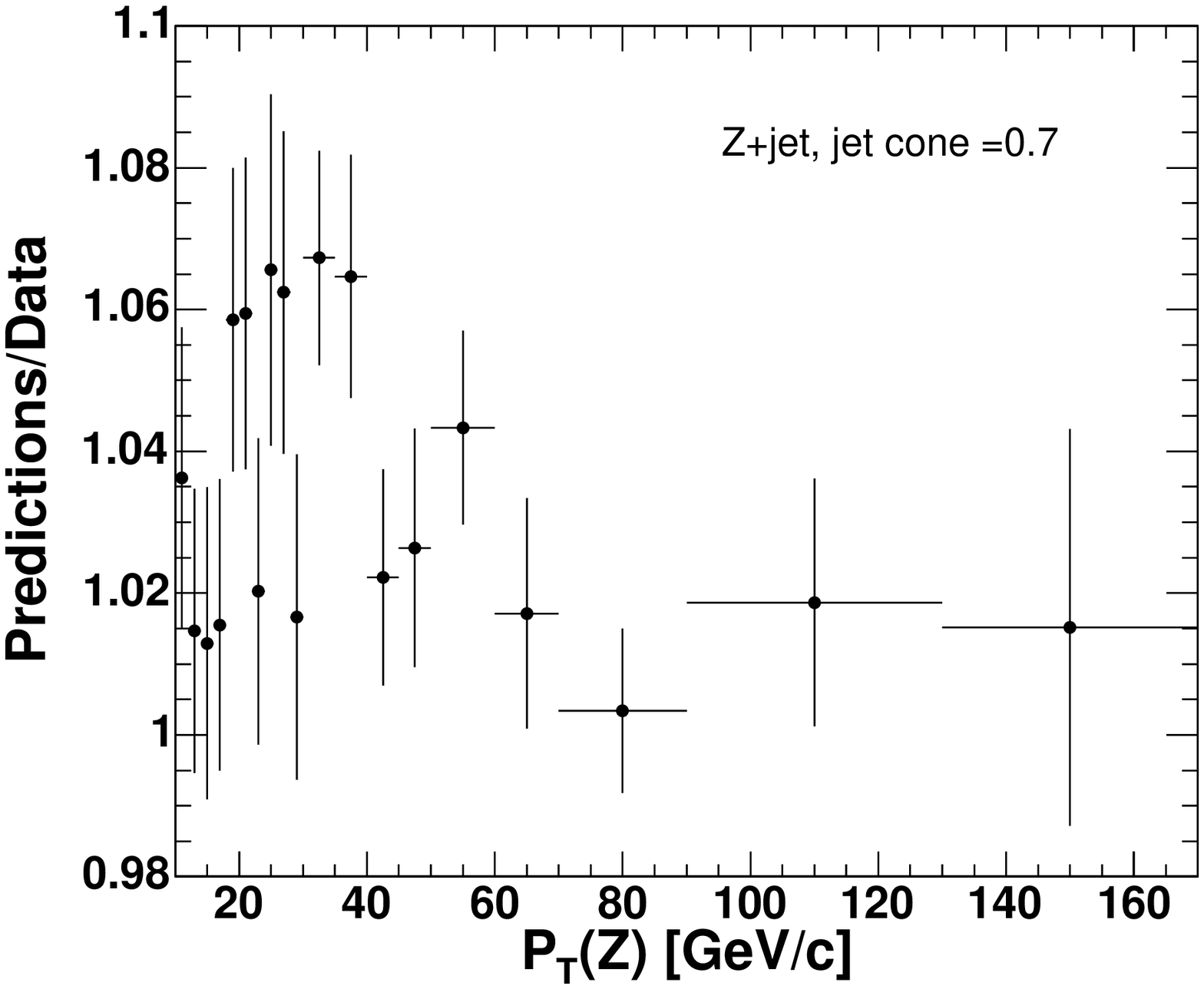}}
\caption{ a.) The average \pt-balance as a function of \pt($Z$). b.)
The ratio of predicted and measured distributions in \pt-balance. The
predicted distribution is for the combination of quark and gluon jets
given by {\sc pythia}. The jets are clustered using a cone radius of
R=0.7. The \pt-balance is noticeably distorted by the cut-off on the
minimum \pt~of the leading jet and a finite jet energy resolution in
events with \pt$(Z)$ $<$ 25 \GeVc. These events will not be used for
the study of the predicted jet energy.}
\label{fig:balance_cone07}
\end{figure}

\begin{figure}[h]
\centering
\subfloat[][]{\includegraphics[angle=0,width=0.49\textwidth]{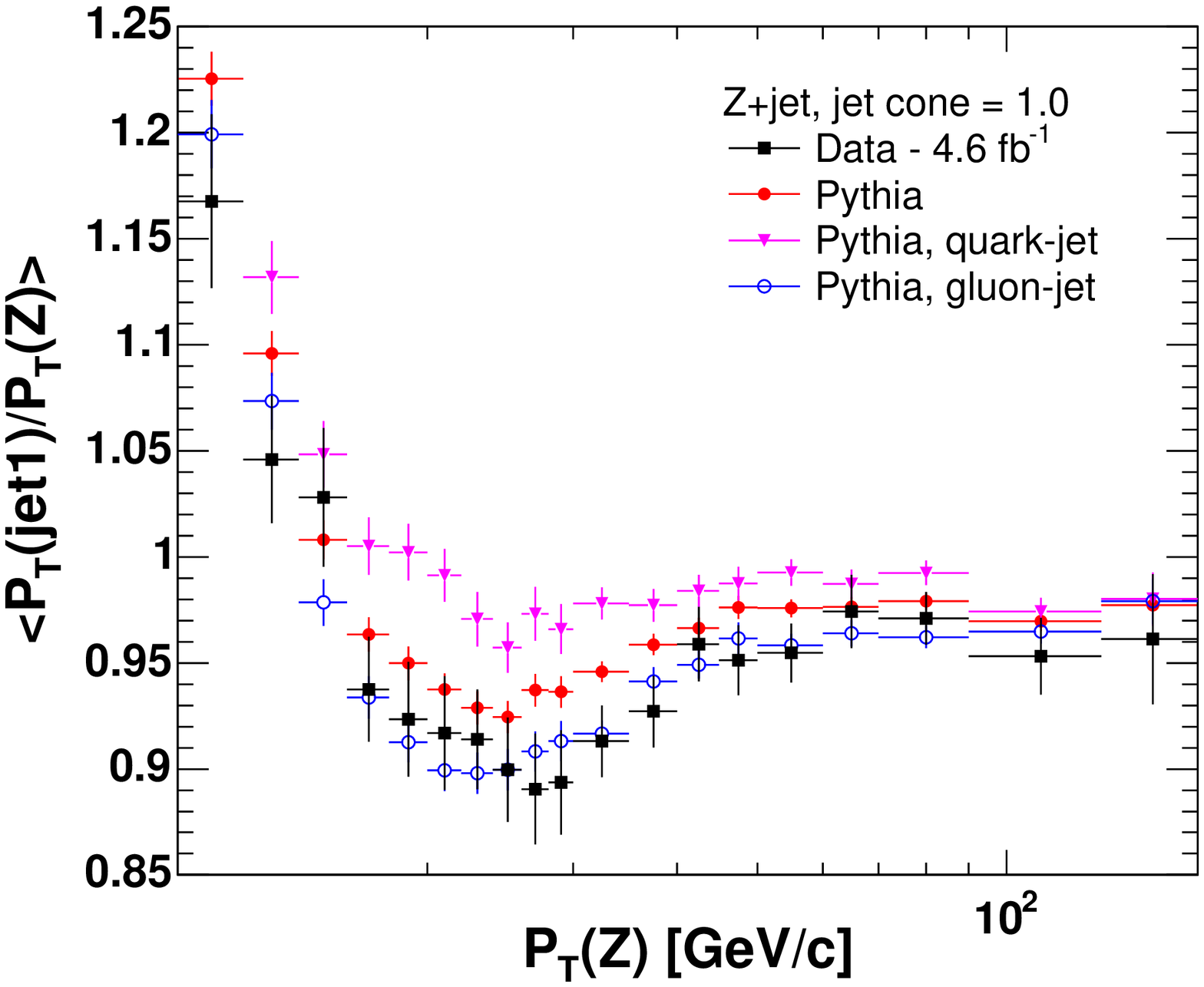}}
\subfloat[][]{\includegraphics[angle=0,width=0.49\textwidth]{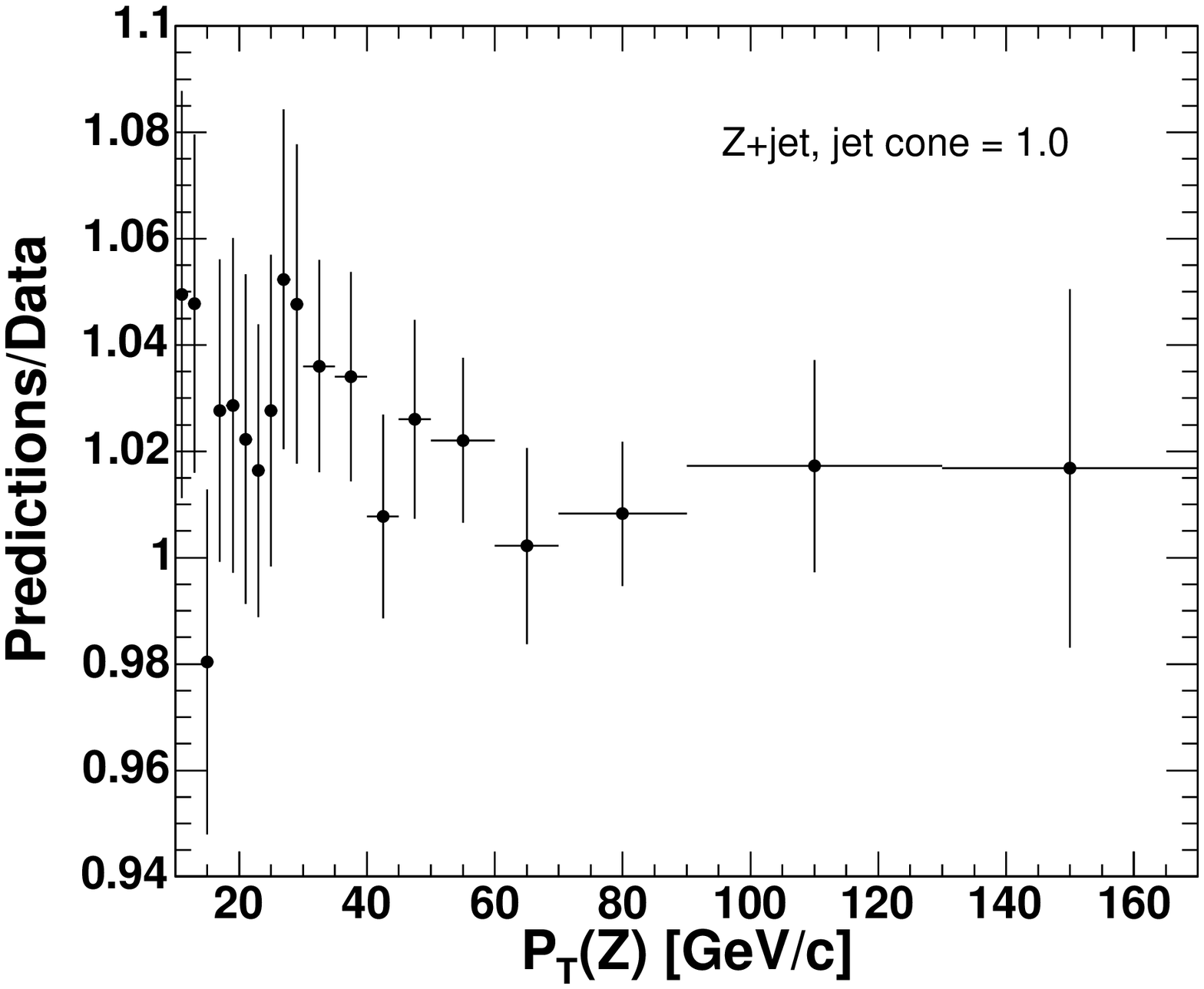}}
\caption{ a.) The average \pt-balance as a function of \pt($Z$). b.)
The ratio of predicted and measured distributions in \pt-balance. The
predicted distribution is for the combination of quark and gluon jets
given by {\sc pythia}. The jets are clustered using a cone radius of
R=1.0. The \pt-balance is noticeably distorted by the cut-off on the
minimum \pt~of the leading jet and a finite jet energy resolution in
events with \pt$(Z)$ $<$ 25 \GeVc. These events will not be used for
the study of the predicted jet energy.}
\label{fig:balance_cone10}
\end{figure}

 A quark jet deposits more energy in the calorimeter system on
average than a gluon jet with the same true momentum. The difference
is caused by the non-linear response of the calorimeter to single
particles and the different multiplicities of hadrons. The predicted
\pt-balances are presented as a function of \pt$(Z)$ for quark and
gluon jets and for data in Figs.~\ref{fig:balance_cone04},
\ref{fig:balance_cone07}, and~\ref{fig:balance_cone10} for jet cone
radii of 0.4, 0.7, and 1.0. The \pt-balance for quark jets is
significantly different than that for gluon jets; it is consequently
essential to check that the mixture of quark and gluon jets is
predicted accurately by {\sc pythia}.

\par The fraction of quark and gluon jets in the $Z$-jet sample is
largely driven by the parton distribution functions of the colliding
particles (e.g. $p\bar{p}$ at the Tevatron and $pp$ at the LHC) and
the matrix elements of $q\bar{q}\goes Zg$ and $qg\goes Zq$ tree-level
diagrams. The \pt-balance observed in data is different than that from
{\sc pythia} simulations as shown in Figs.~\ref{fig:balance_cone04},
\ref{fig:balance_cone07}, and~\ref{fig:balance_cone10}.
\par We test that the discrepancy between data and predictions in the
\pt-balance is not caused by an incorrectly modeled fraction of quark
and gluon jets using two methods. We compare rapidity distributions
for $Z$+jet events in Section~\ref{rapid_Zj} to validate the relative
contributions from $qg\goes Zq$ and $q\bar{q}\goes Zg$ LO diagrams in
{\sc pythia} and {\sc alpgen}. Having the rapidity distributions for
data, {\sc alpgen}, and {\sc pythia} in good agreement, we further
test the prediction from {\sc pythia} alone by looking at the number
of tracks inside the jet cone (see Section~\ref{N_tracks}).

\par The quark-gluon mixture is sensitive to PDFs, choice of
renormalization and factorization scales, and LO matrix element for
the $Z$ + 1-parton final state. The impact from these will be reported
later in Sections~\ref{Py_Alp}, \ref{PDFs}, \ref{ISR},
and~\ref{Q2_Alp}.

%
\subsection{Kinematic properties of $Z$+jet events}
\label{rapid_Zj}
\par
\begin{figure}[h]
\centering
\subfloat[][]{\includegraphics[angle=0,width=0.49\textwidth]{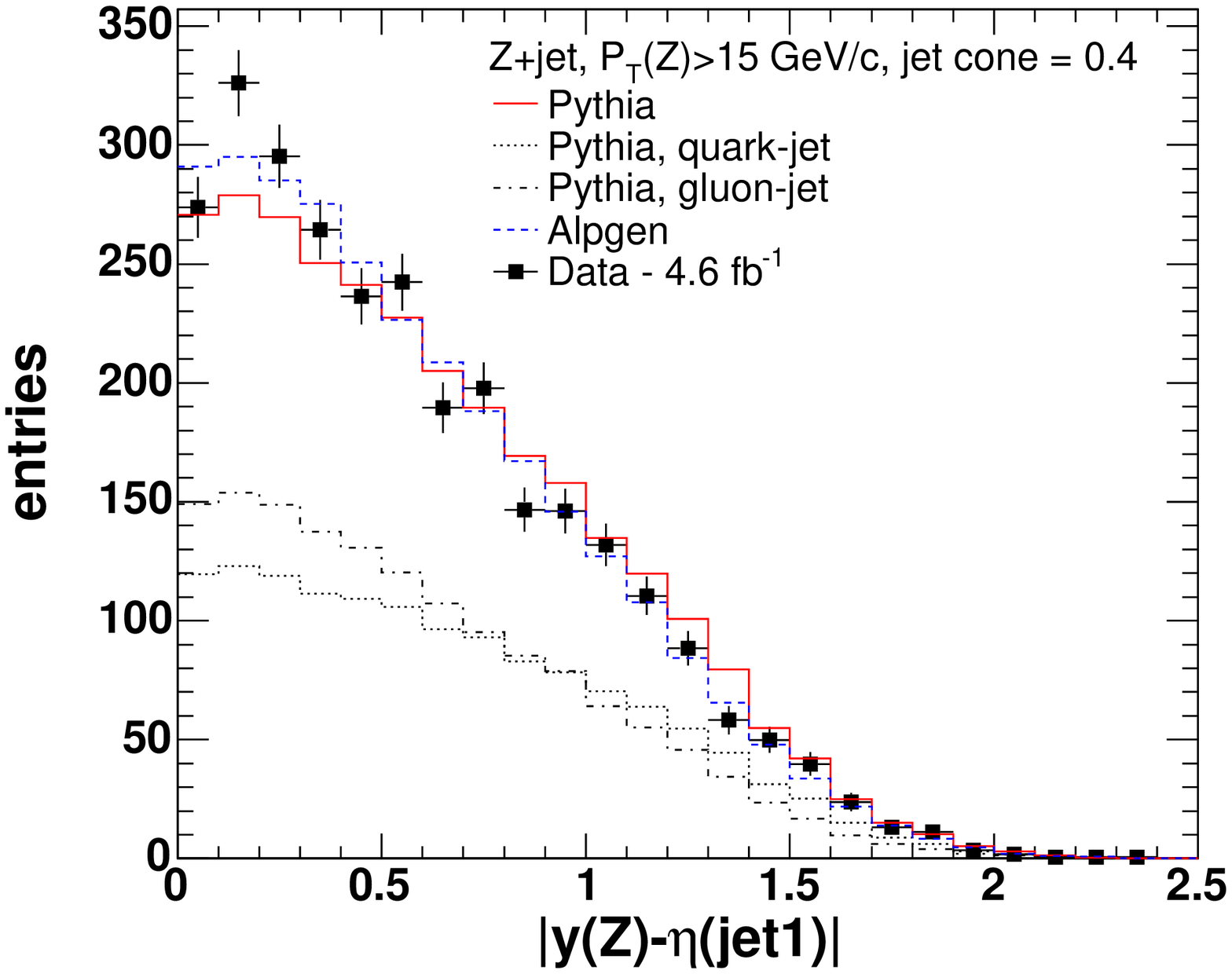}}
\subfloat[][]{\includegraphics[angle=0,width=0.49\textwidth]{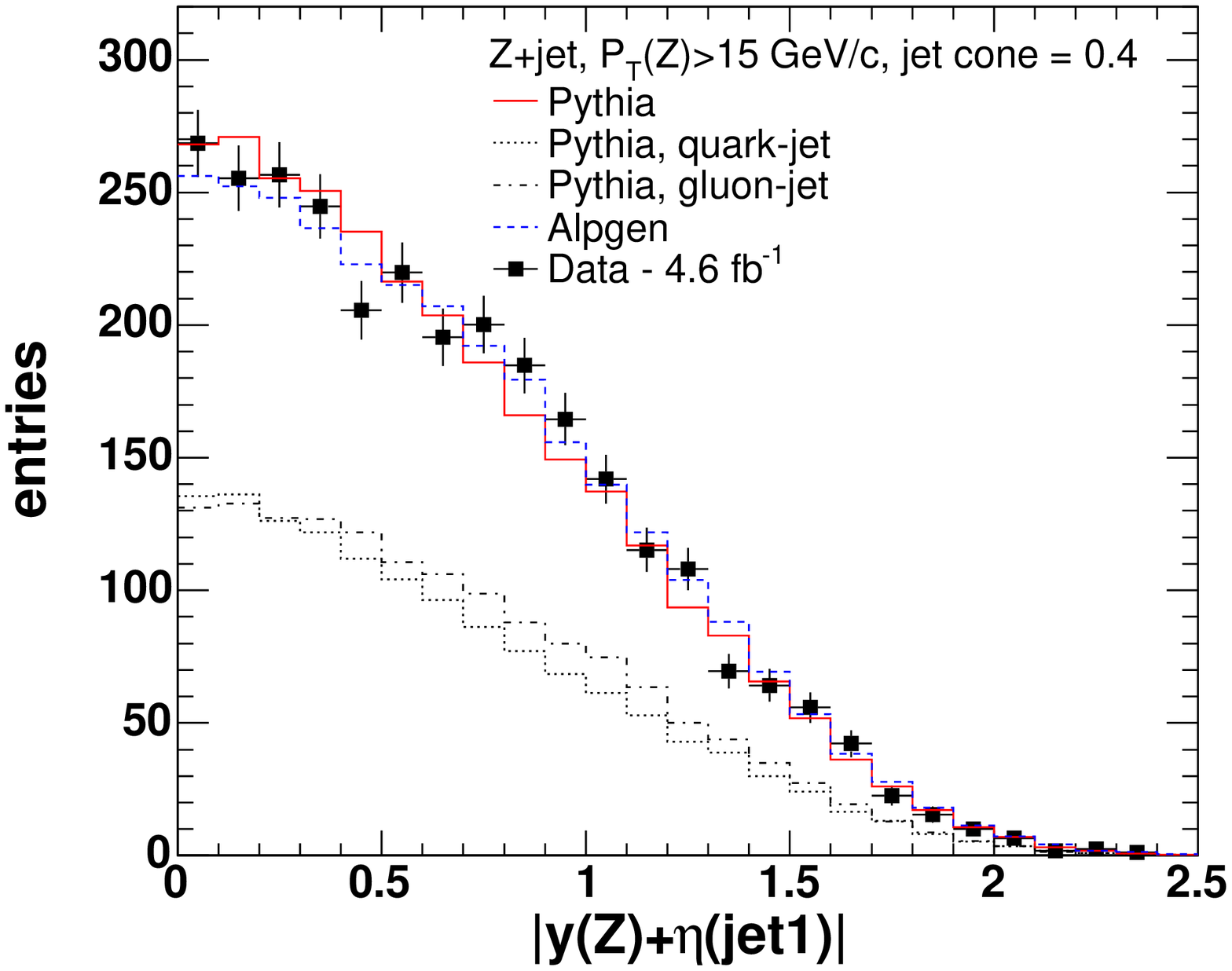}}
\caption{The rapidity distributions for the $Z$+jet system. The jet
clustering is performed with a cone of R=0.4. Figure (a) is difference
of the $Z$ and jet rapidities; Figure (b) is the sum.}
\label{fig:yZ_cone04}
\end{figure}

 The kinematic properties of $Z$+jet events provide an indirect
test of the quark-gluon composition of the leading jet. SM predictions
are studied using the distributions of sum and difference of
rapidities of a $Z$ boson and the leading jet, $|y(Z)+\eta(jet1)|$ and
$|y(Z)-\eta(jet1)|$, respectively (see Fig.~\ref{fig:yZ_cone04}). We
require \pt$(Z)$ $>$ 15~\GeVc to avoid very soft jets.  We observe
good agreement between data and the predictions (both {\sc alpgen} and
{\sc pythia}). The distributions are slightly different for {\sc
alpgen} and {\sc pythia} due to the different methods of computing MEs
for the tree-level processes.

\subsection{Charged particle multiplicity in jets}
\label{N_tracks}
\par
\begin{figure}[h]
\centering
\subfloat[][]{\includegraphics[angle=0,width=0.49\textwidth]{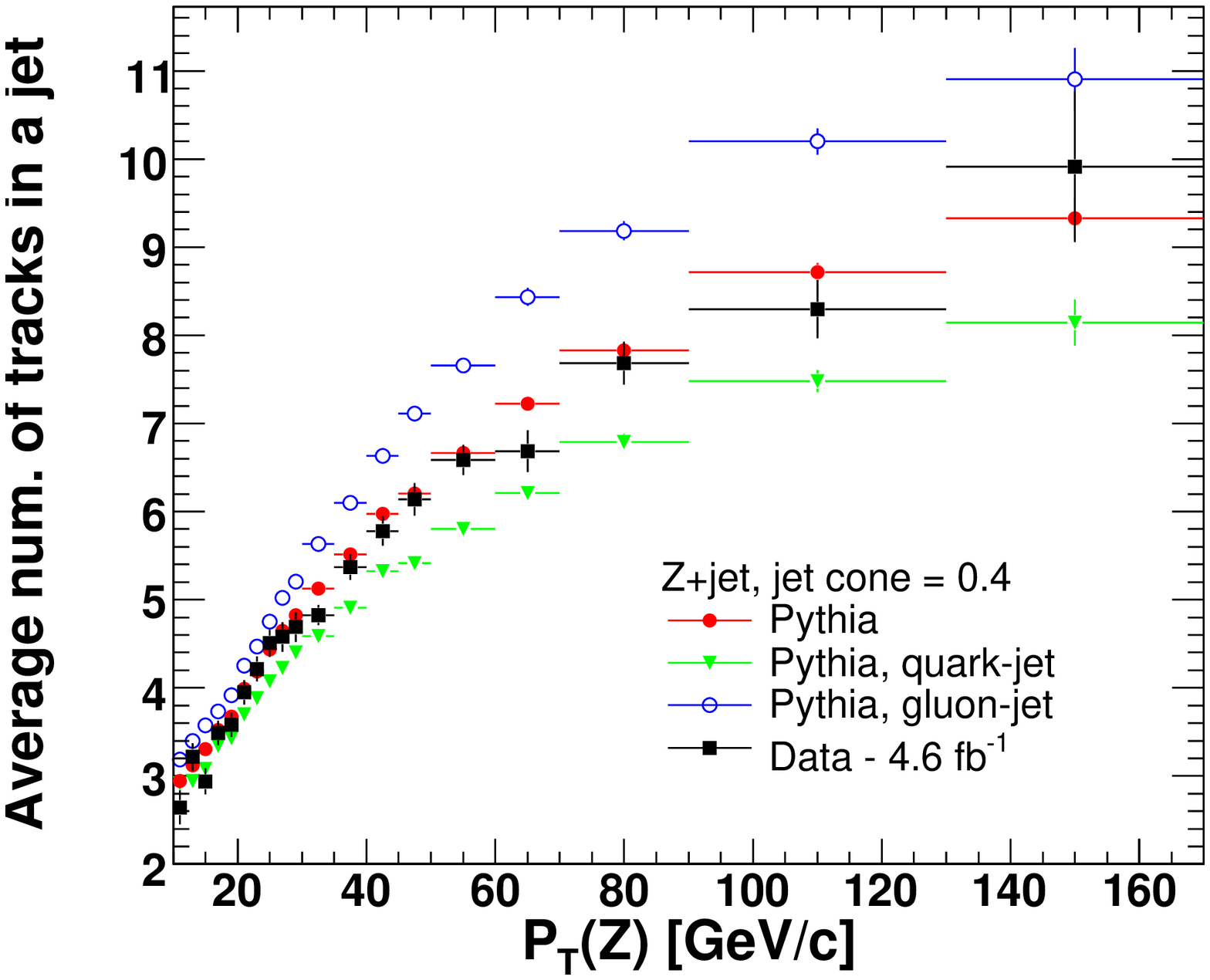}}
\subfloat[][]{\includegraphics[angle=0,width=0.49\textwidth]{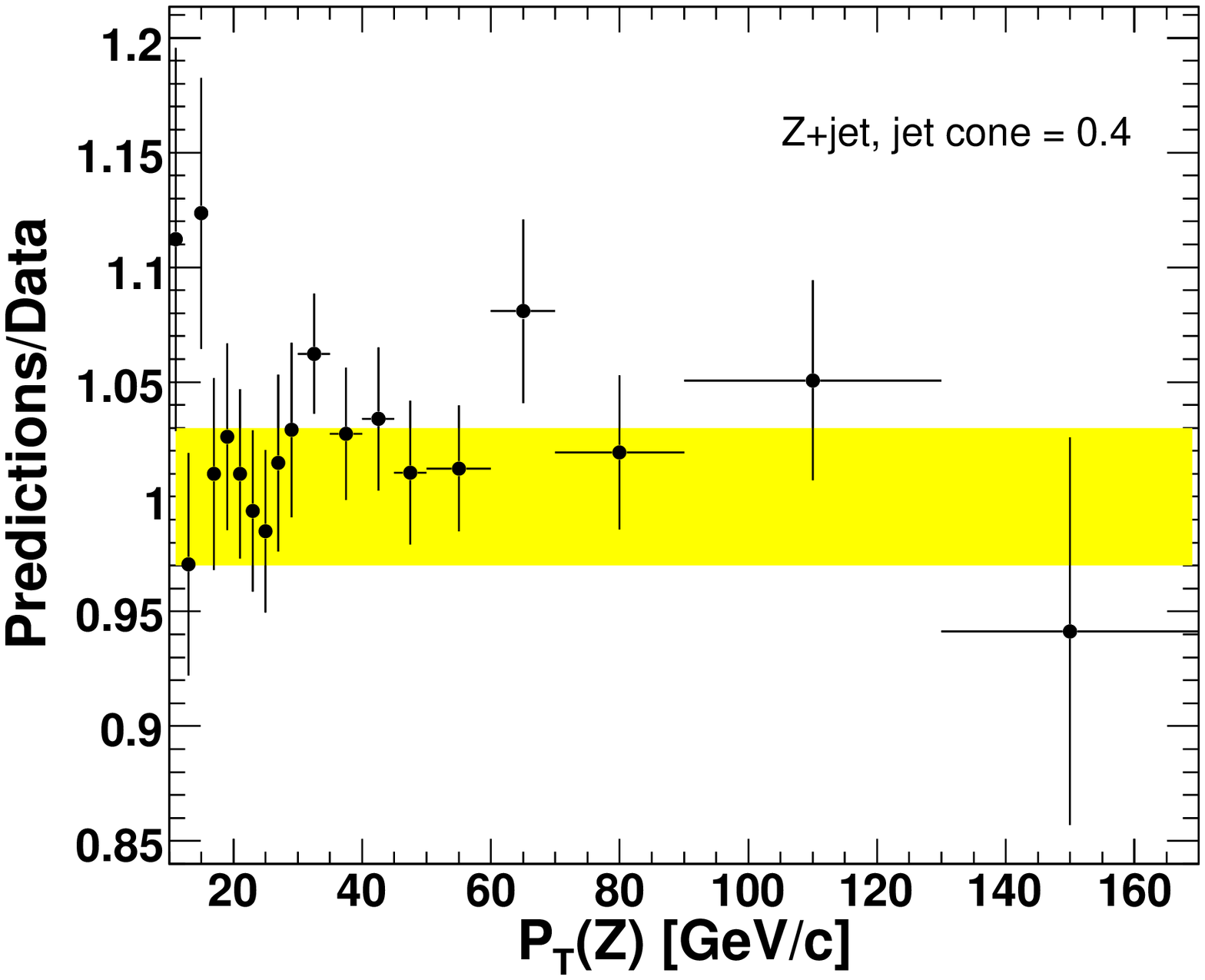}}
\caption{ a) The average number of tracks within a jet cone of radius
of R=0.4 as a function of \pt($Z$). b) The ratio of the predicted
number of tracks to the measured number in data versus \pt($Z$). The
solid band represents a 3\% uncertainty on the predicted tracking
efficiency~\cite{track_efficiency}.}
\label{fig:Ntrk_c04}
\end{figure}

 We perform a direct test of the quark-gluon composition of the
observed jets by using the number of tracks observed within the cone
of the leading jet. The tracks are required to originate from the same
vertex as the lepton pair forming a $Z$ boson, $|z_0 - z_{track}|$ $<$
4 cm and $|d_0(tracks)|$ $<$ 0.02 cm with SVX hits (0.2 cm without
SVX). Also we require a good track quality for the tracks, namely
$\chi^2$ for the track fit divided by the number of degrees of freedom
should be less than 6. The transverse momentum of the tracks is
required to be greater than 0.3~\GeVc. The number of tracks is
different for quark and gluon jets (see Fig.~\ref{fig:Ntrk_c04}).
Overall, the observed events are in good agreement with the SM
predictions ({\sc pythia}).
\par Both tests of the quark-gluon composition of jets have
demonstrated that {\sc pythia} simulations describes the jet
composition accurately. From now we proceed with investigating the
other sources of uncertainties, which contribute to the discrepancy in
\pt-balance between data and {\sc pythia}
(Figs.~\ref{fig:balance_cone04}, \ref{fig:balance_cone07},
and~\ref{fig:balance_cone10}).

\section{Uncertainty due to calculation of the matrix elements and the 
jet-parton matching scheme}
\label{Py_Alp}
\par Calculation of matrix elements (MEs) for $Z$+0 partons and $Z$+1
parton is different in {\sc alpgen+pythia} and stand-alone {\sc
pythia}. In addition, {\sc alpgen+pythia} simulation uses a veto
algorithm to avoid double counting of jets produced by matrix elements
and radiation from the parton shower~\cite{MLM_matching}. The
difference in methods affects the \pt-balance obtained using the two
event generators. To estimate the difference we calculate the
\pt-balance as a function of \pt$(Z)$ in both the {\sc alpgen} and
{\sc pythia} samples similarly to that shown in
Figs.~\ref{fig:balance_cone04}, \ref{fig:balance_cone07},
and~\ref{fig:balance_cone10}. Then we calculate the ratio between the
distributions as a function of \pt$(Z)$. The mean value of the ratio
for events with \pt$(Z)$ $>$ 25 \GeVc is taken as the uncertainty due
to calculation of MEs and jet-parton matching. The resulting
uncertainty is found to be about 1\% as shown in
Table~\ref{table:systematics}.

\section{PDF uncertainties}
\label{PDFs}
\par We test the sensitivity of the \pt-balance to the choice of the
PDF set used to generate events~\cite{Pumplin:2002vw}. The default PDF
set, CTEQ5L, is a single set of functions and it does not contain
error functions~\cite{Hessian_method}. We estimate the PDF-related
systematic uncertainty using the Hessian method; we take the CTEQ6M
set, which includes 40 eigenvector error functions in addition to the
central value. We re-weight the existing events with the parton
densities provided by the CTEQ6M set. For each error set we calculate
the difference in \pt-balance relative to the central value given by
CTEQ6M. The sum in quadratures of the 40 variations in \pt-balance
results in a negligibly small value (less that 0.1\%) of the
uncertainty on the predicted balance.

%
\section{ISR uncertainties}
\label{ISR}
\par In the Monte Carlo sample of inclusive $Z$ events generated using
{\sc pythia}, most of the jets arise from the ISR parton shower model
(the underlying event model also contributes to production of soft
jets), which depends upon several parameters that are not tightly
constrained by data. As a result, the parameters of ISR affect the
observed \pt-balance in $Z$+jet events.


\par To produce the systematics samples we altered the ISR settings
used in {\sc pythia}, parameters PARP(61) and PARP(64), similarly to
the procedure described in~\cite{top_mass_jes}. The parameter PARP(61)
defines $\Lambda_{\rm QCD}$ used in running of $\alpha_s$ in
space-like parton showers. By default, the value of $\Lambda_{\rm
QCD}$ is chosen according to the PDF parameterizations. Parameter
PARP(64) is also used in the calculation of $\alpha_s$ and parton
distributions as a multiplier for the squared transverse momentum
evolution scale, $k^2_{\perp}$.

\par The variations in \pt-balance calculated for the systematics
samples relative to the default are used to estimate the ISR
uncertainty. The resulting value is on the order of 1\%.

\section{Uncertainty due to renormalization and factorization scales}
\label{Q2_Alp}
\par Predictions for $Z$+jet production are sensitive to the choice of
renormalization and factorization scales. The accuracy of LO
calculations is often estimated by varying renormalization and
factorization scales (NLO calculations demonstrate weaker dependence
on the scales)~\cite{PhysRevD.40.2888}. The scales impact calculation
of the LO matrix elements for events with a $Z$ boson and N
partons. We exploit {\sc alpgen} to generate events with altered
renormalization and factorization scales. The scales are always kept
the same by {\sc alpgen}, with a default value $Q_0=\sqrt{M^2_Z+\Sigma
p_T^2(jet)}$. We produce two ``systematic'' samples with $Q_0$ being
multiplied by 0.5 or 2.0. The choice of the scales impacts the
predicted \pt-balance by up to 1-2\% as recorded in
Table~\ref{table:systematics}.
\par The effect due to variation of renormalization and factorization
scales in {\sc alpgen} is similar to that caused by variation of the
ISR parameters in {\sc pythia} (see Sec.~\ref{ISR}). The stand-alone
{\sc pythia} method of calculations for the inclusive production of
$Z$ bosons uses an ISR parton shower to estimate the LO matrix element
for $Z$ plus one parton. Both methods deal with the same computational
uncertainty caused by limited accuracy of the LO calculations used in
the two MC event generators so that the variation of scales results in
similar uncertainties for the both generators. We use the variation in
\pt-balance estimated with {\sc alpgen} and not that obtained with
{\sc pythia} to avoid double-counting the effect. The resulting effect
on the predicted \pt-balance is about 1-2\% as recorded in
Table~\ref{table:systematics}.
%

\section{Characteristics of out-of-cone radiation}
\label{OOC_rad}
\par An understanding of the energy flow outside of the cone of the
leading jet is essential for interpreting the measurement of
\pt-balance in $Z$-jet events. Raw calorimeter energy summed in annuli
outside of the jet cone is not linearly proportional to the true
out-of-cone energy.

\par We exploit correlations between \pt-balance and properties of the
sub-leading jet such as \pt(jet2) and $\Delta\phi(jet1-jet2)$.
Multiple \ppbar~interactions produce additional soft jets that are
unrelated to the jets recoiling against the $Z$-boson. The presence of
additional interactions might bias the correlation between the
\pt-balance and the properties of the sub-leading jet. In the method
described in this section we require all events to have exactly one
primary vertex, {\it i.e.} one interaction. The requirement to have
exactly one primary vertex was not applied before since we did not
have to measure jet clusters with \pt~softer than 8~\GeVc.
\par
\begin{figure}[h]
\centering
\subfloat[][]{\includegraphics[angle=0,width=0.49\textwidth]{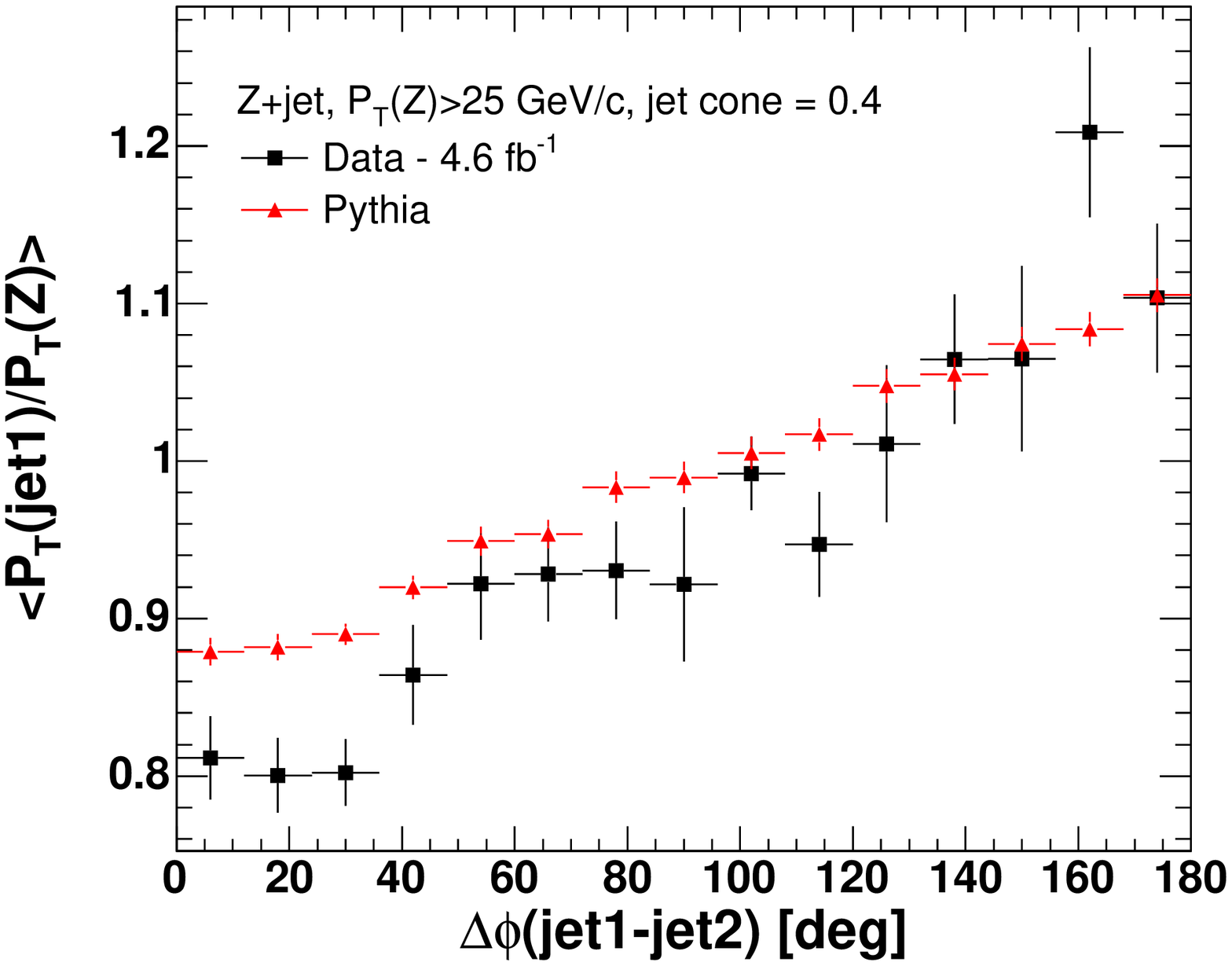}}
\subfloat[][]{\includegraphics[angle=0,width=0.49\textwidth]{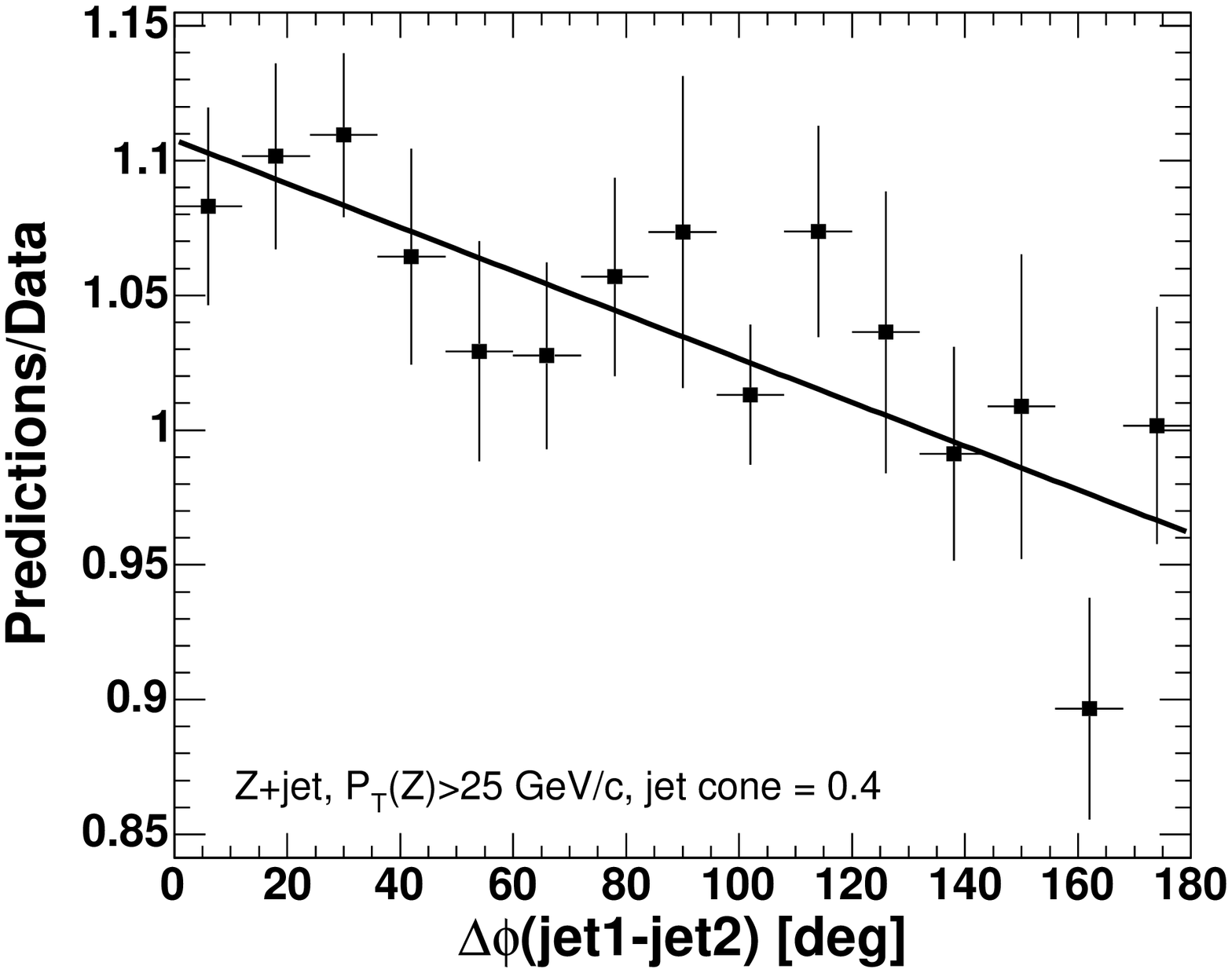}}
\caption{a.) A comparison of the measured (square markers) and
predicted (triangle markers) \pt-balance as a function of
$\Delta\phi(jet1-jet2)$ for jets of R=0.4 cone size. The predicted
balance is obtained with {\sc pythia}. The events are required to have
an only one interaction per event. b.) The fit of the ratio to a line
results in $\chi^2/NDF$ = 10.4/14 and slope = -8.12$\cdot10^{-4}$
$\pm$ 1.96$\cdot10^{-4}$, as could be explained by an inadequate
modeling of large-angle parton radiation.}
\label{fig:j12_phi_cone04}
\end{figure}

\begin{figure}[h]
\centering
\subfloat[][]{\includegraphics[angle=0,width=0.49\textwidth]{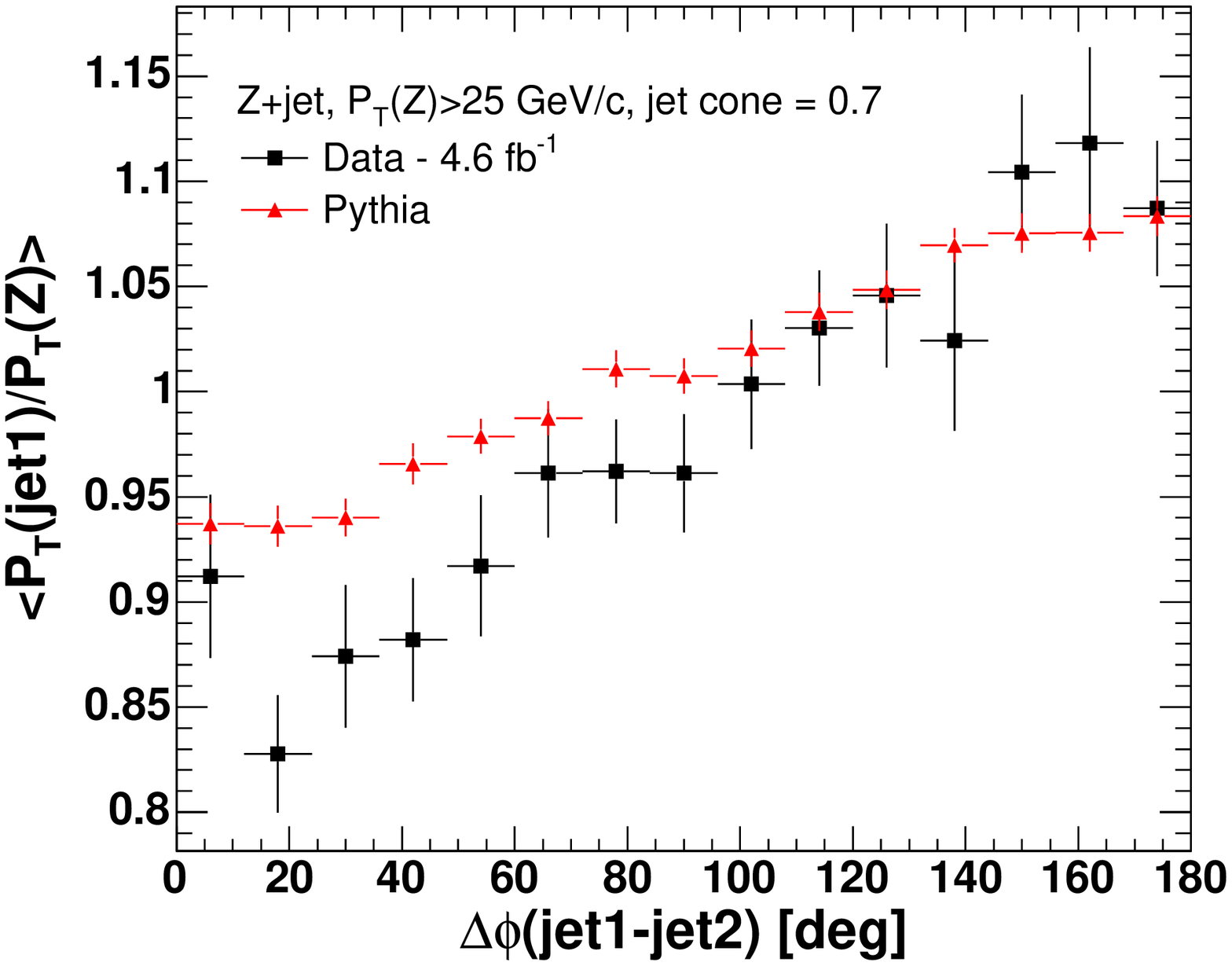}}
\subfloat[][]{\includegraphics[angle=0,width=0.49\textwidth]{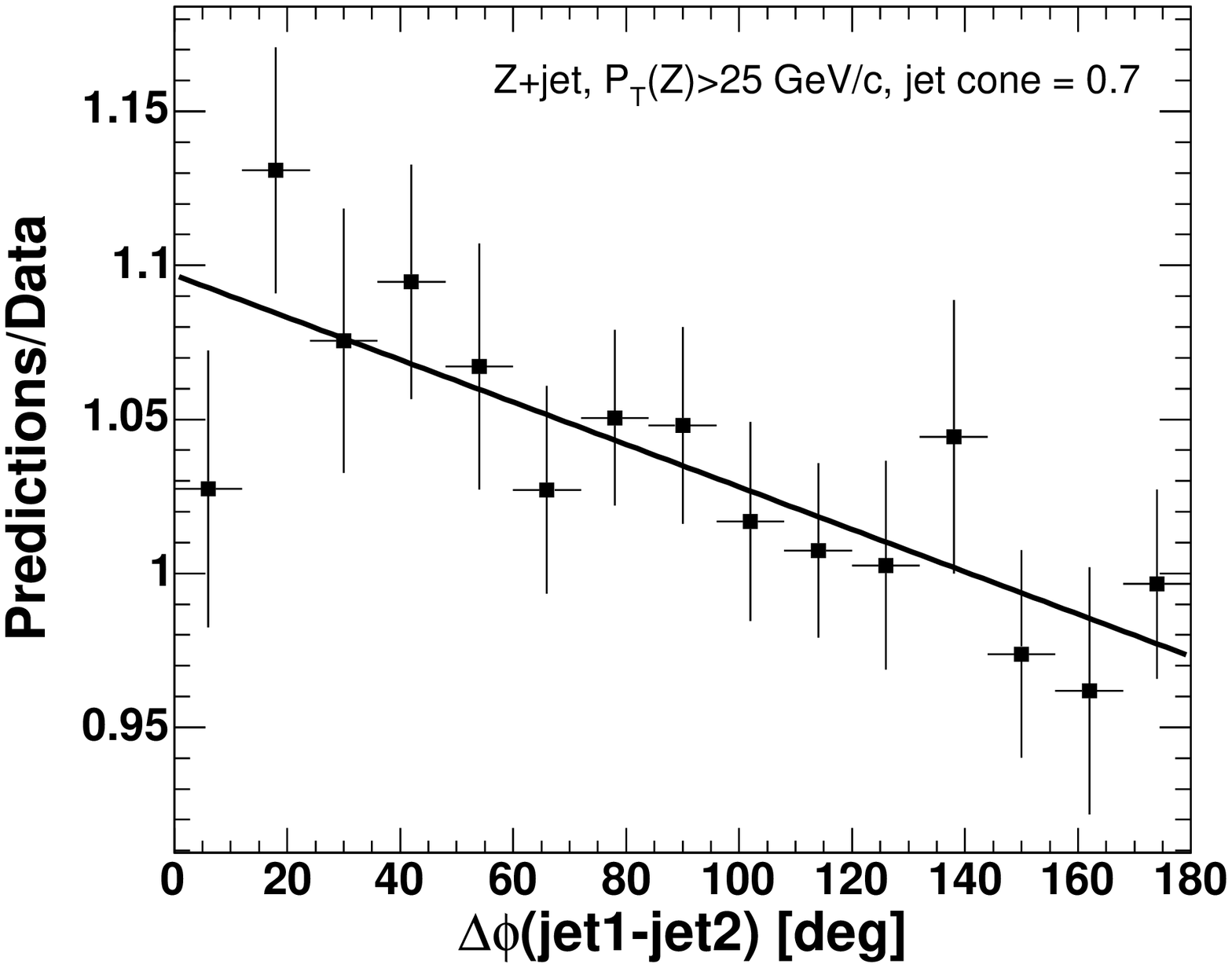}}
\caption{a.) A comparison of the measured (square markers) and
predicted (triangle markers) \pt-balance as a function of
$\Delta\phi(jet1-jet2)$ for jets of R=0.7 cone size. The predicted
balance is obtained with {\sc pythia} The events are required to have
an only one interaction per event. b.) The fit of the ratio to a line
results in $\chi^2/NDF$ = 7.0/14 and slope = -6.9$\cdot10^{-4}$ $\pm$
1.9$\cdot10^{-4}$. }
\label{fig:j12_phi_cone07}
\end{figure}

\begin{figure}[h]
\centering
\subfloat[][]{\includegraphics[angle=0,width=0.49\textwidth]{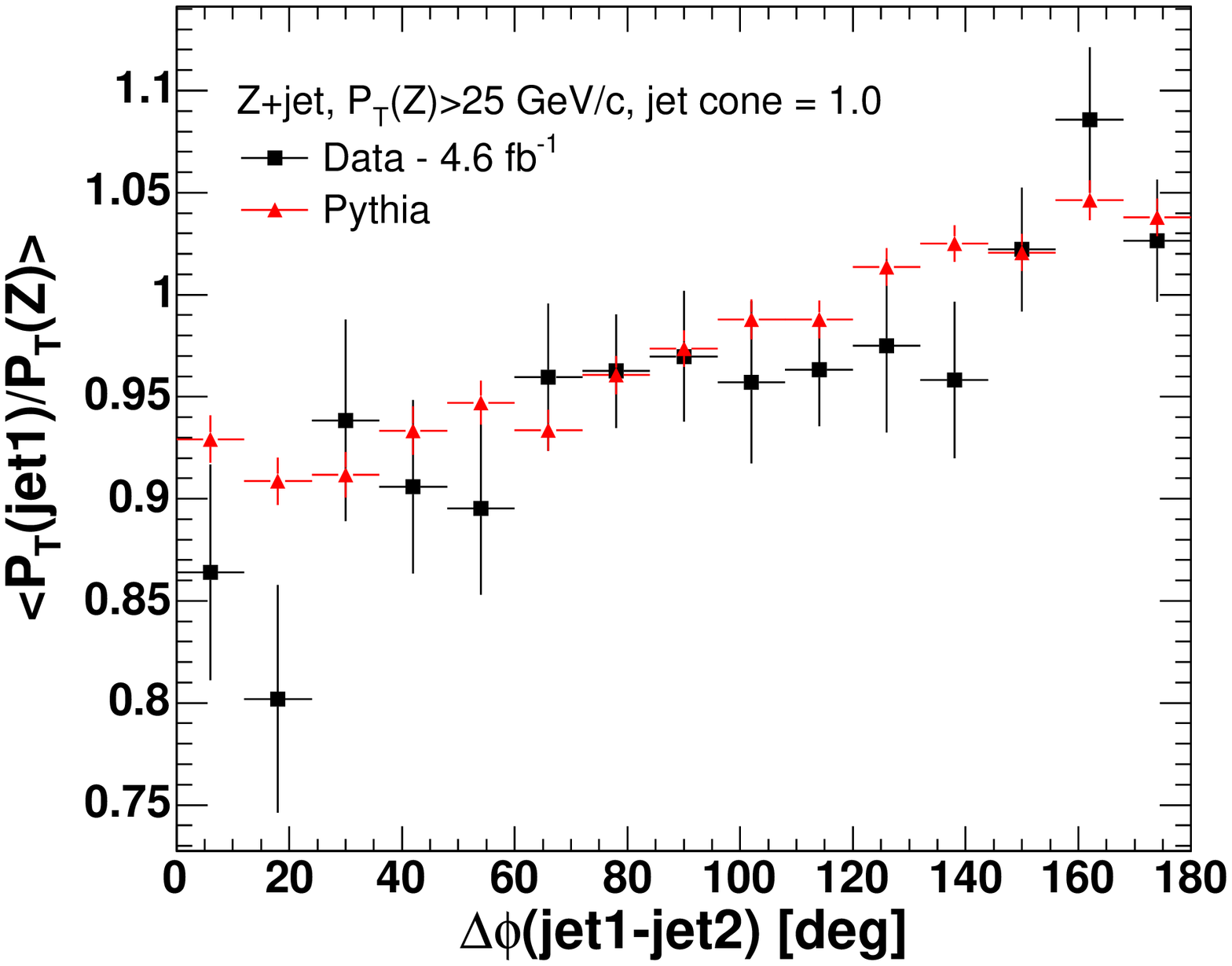}}
\subfloat[][]{\includegraphics[angle=0,width=0.49\textwidth]{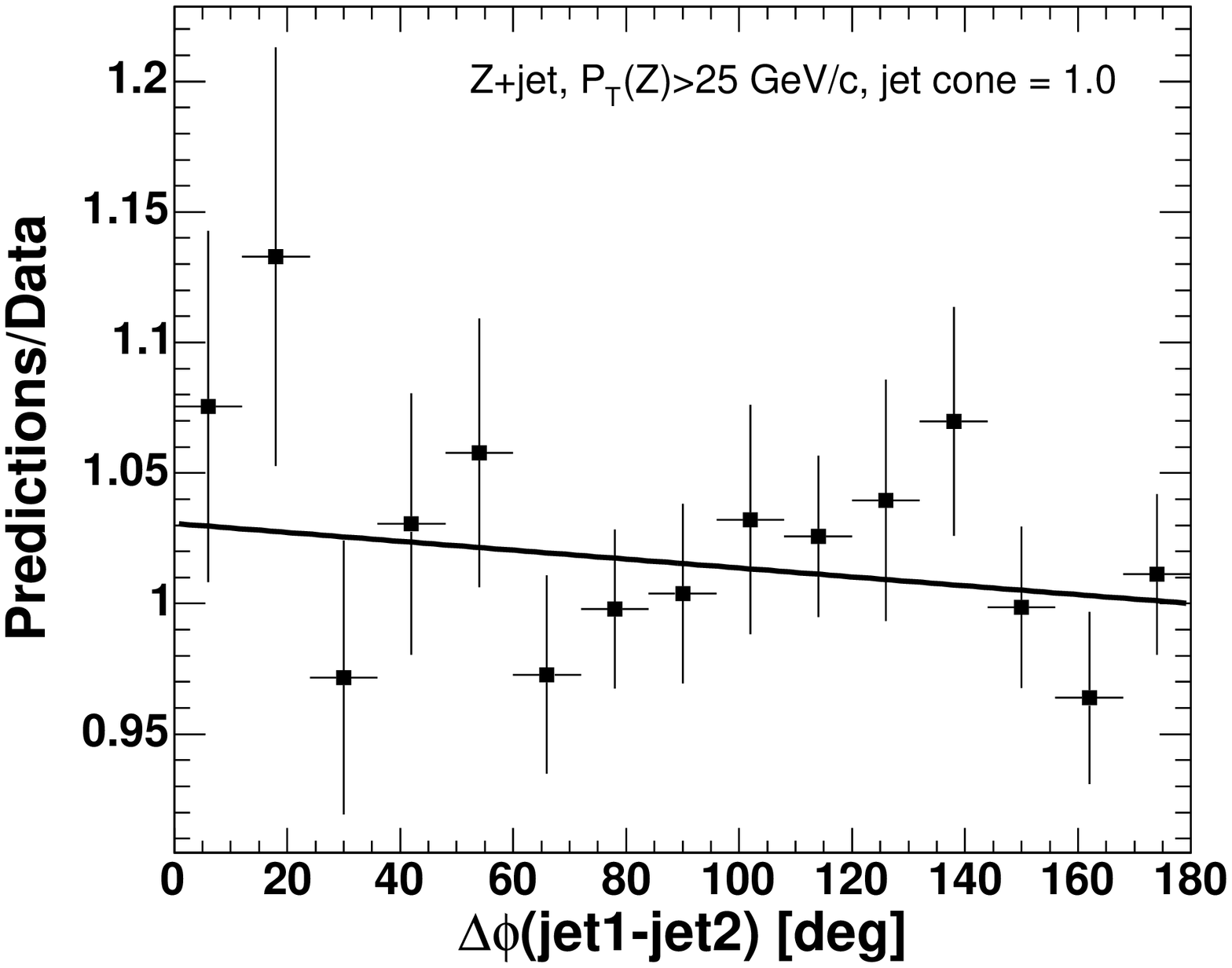}}
\caption{a.)  A comparison of the measured (square markers) and
predicted (triangle markers) \pt-balance as a function of
$\Delta\phi(jet1-jet2)$ for jets of R=1.0 cone size. The predicted
balance is obtained with {\sc pythia}. The events are required to have
an only one interaction per event. b.) The fit of the ratio to a line
results in $\chi^2/NDF$ = 10.2/14 and slope = -1.7$\cdot10^{-4}$ $\pm$
2.2$\cdot10^{-4}$. }
\label{fig:j12_phi_cone10}
\end{figure}

 We measure the dependence of the \pt-balance on the azimuthal
angle between the leading jet ($jet1$) and the sub-leading one
($jet2$), $\Delta\phi(jet1-jet2)$, for events with \pt$(Z)$ $>$
25~\GeVc (see Figs.~\ref{fig:j12_phi_cone04},
\ref{fig:j12_phi_cone07}, and \ref{fig:j12_phi_cone10}). The observed
dependence of the \pt-balance on the angular separation between the
1st and the 2nd jets is sensitive to the jet cone size. As expected
the largest discrepancy between data and predictions is observed with
the smallest jet cone size of R=0.4 (see
Fig.~\ref{fig:j12_phi_cone04}), for which the particle-jet-energy to
parton-jet-energy correction is at maximum. Larger jet cone sizes give
result in better agreement between data and {\sc pythia} simulations.
Two cone-0.4 jets can be reconstructed as one cone-1.0 jet. As a
result the large-angle radiation most likely falls within the jet cone
of R=1.0.
\par The positive correlation between the \pt-balance and
$\Delta\phi(jet1-jet2)$ shows that the 2nd jet is often caused by the
parton radiation from the leading jet as the magnitude of the
correlation is proportional to the rate of the large-angle parton
radiation. A negative slope of the ratio between data and predictions
(e.g. see Fig.~\ref{fig:j12_phi_cone04}(b)) indicates that the data
exhibit more large-angle parton radiation than the MC simulations.
The deficiency of large-angle parton radiation is also consistent with
the results from~\cite{Abazov:2009pp}.
%

\par We measure the dependence of the \pt-balance on the \pt~of the
second jet, \pt(jet2). The balance as a function of the 2nd jet \pt~is
shown in Figs.~\ref{fig:j2_pt_cone04}, \ref{fig:j2_pt_cone07}, and
\ref{fig:j2_pt_cone10} for different sizes of jet cones. The balance
is most sensitive to the 2nd jet \pt~when the jet cone is 0.4 (see
Fig.~\ref{fig:j2_pt_cone04}). The observed rate of large-angle
out-of-cone radiation is also observed to be higher in data than in
the predictions.

\subsection{Uncertainty due to out-of-cone radiation}
\par
\begin{figure}[h]
\centering
\subfloat[][]{\includegraphics[angle=0,width=0.49\textwidth]{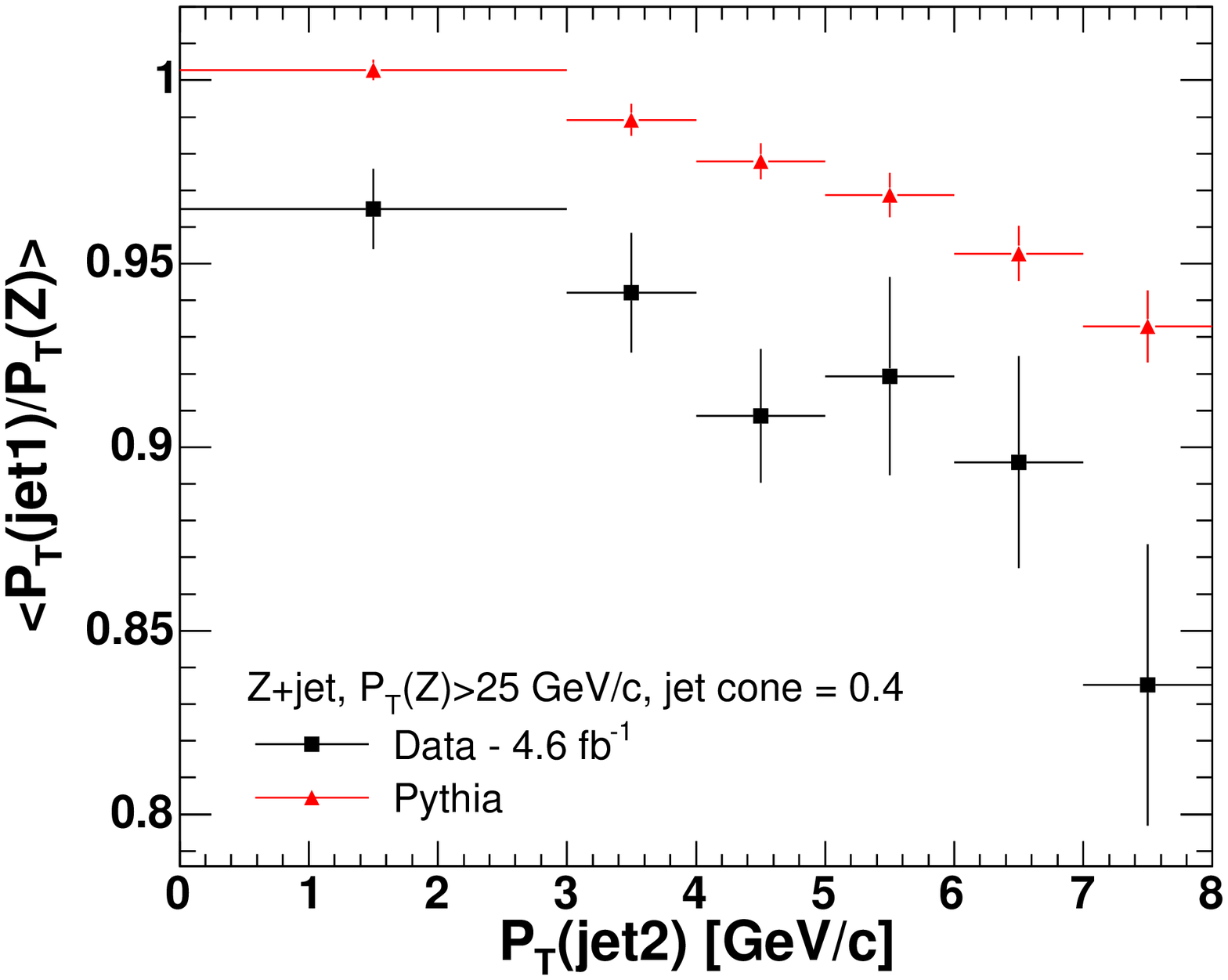}}
\subfloat[][]{\includegraphics[angle=0,width=0.49\textwidth]{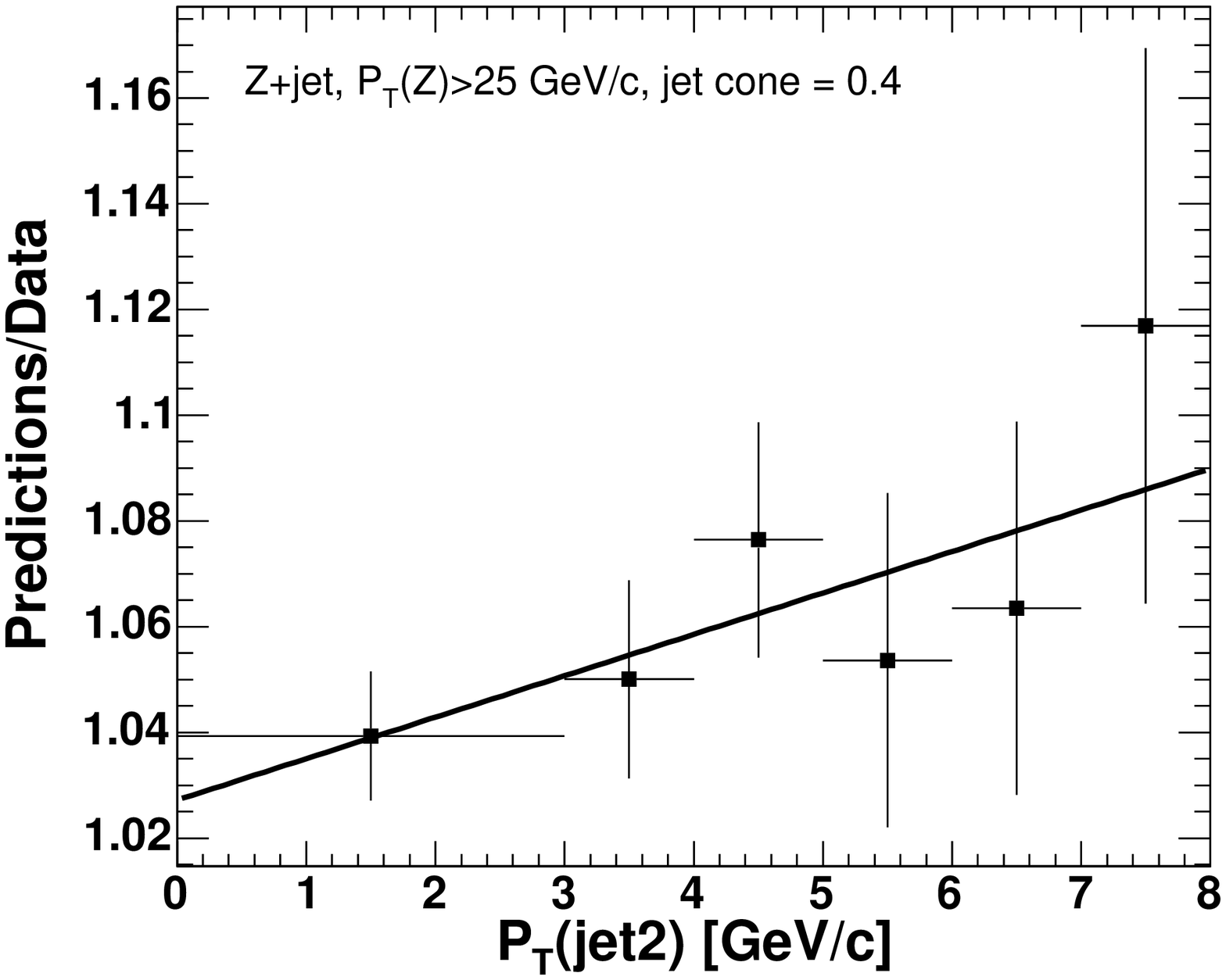}}
\caption{ a.) A comparison of the measured (square makers) and
predicted (triangle markers) \pt-balance as a function of the 2nd jet
\pt~for jets of R=0.4 cone size. The predicted balance is obtained
with {\sc pythia}. The events are required to have only one
interaction per event. b.)  The ratio of predicted to measured
\pt-balance versus the \pt~of the second jet. The linear fit of the
ratio resulted in a slope of 0.78$\pm$0.48 \%/\GeVc.}
\label{fig:j2_pt_cone04}
\end{figure}

\begin{figure}[h]
\centering
\subfloat[][]{\includegraphics[angle=0,width=0.49\textwidth]{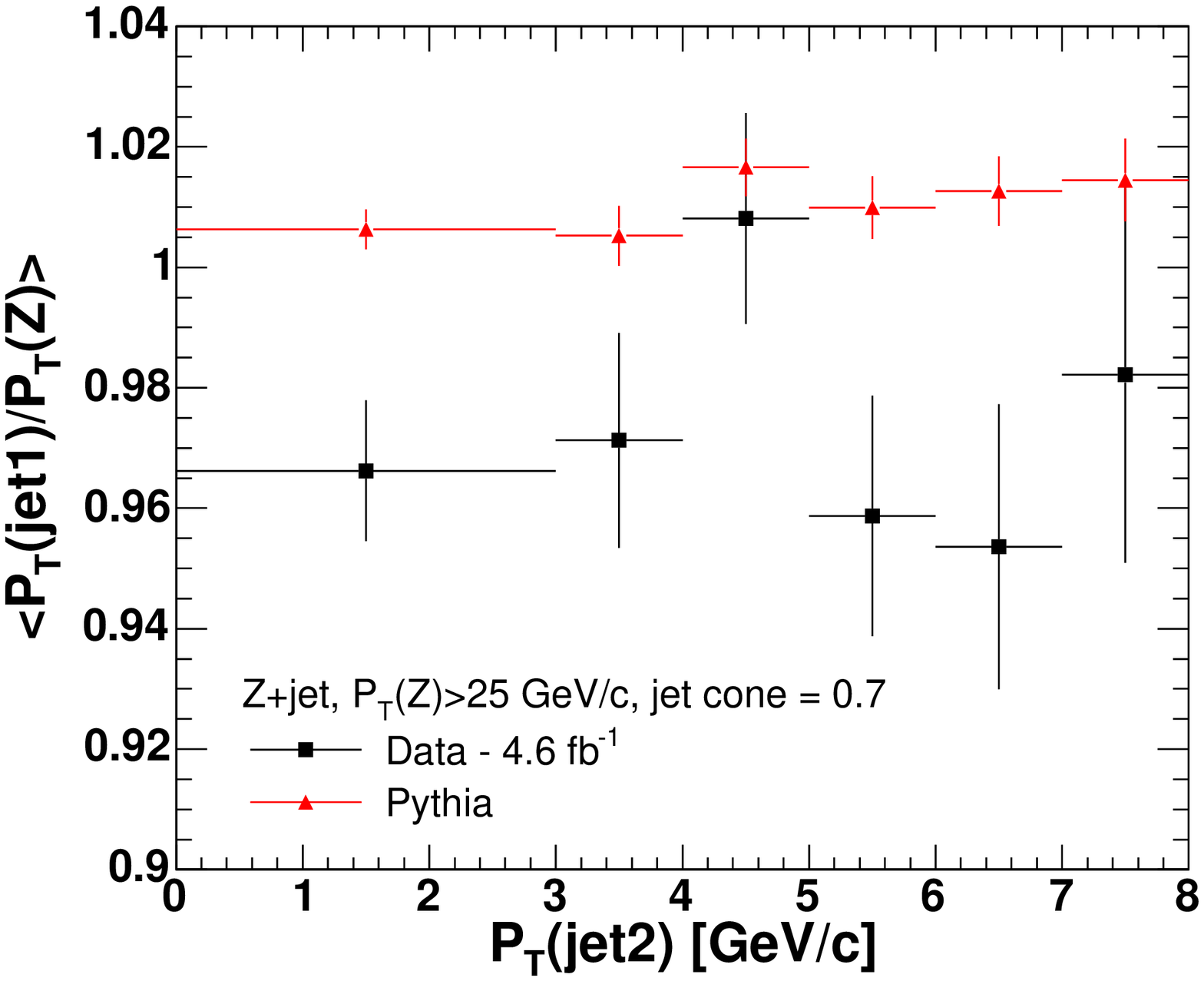}}
\subfloat[][]{\includegraphics[angle=0,width=0.49\textwidth]{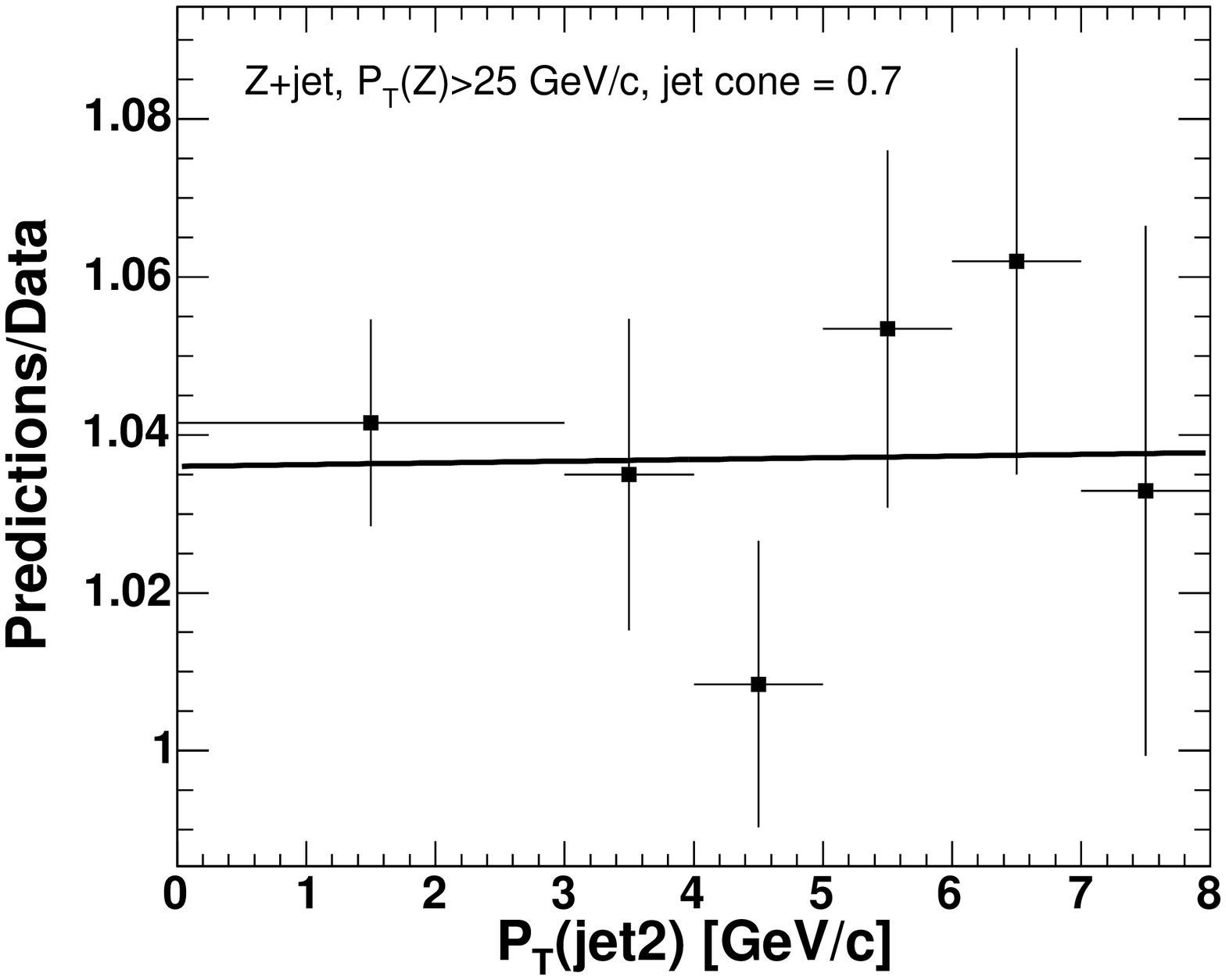}}
\caption{ a.) A comparison of the measured (square markers) and
predicted (triangle markers) \pt-balance as a function of the 2nd jet
\pt~for jets of R=0.7 cone size. The predicted balance is obtained
with {\sc pythia}. The events are required to have only one
interaction per event. b.)  The ratio of predicted to measured
\pt-balance versus the \pt~of the second jet. The linear fit of the
ratio resulted in a slope of 0.02$\pm$0.4 \%/\GeVc.}
\label{fig:j2_pt_cone07}
\end{figure}

\begin{figure}[h]
\centering
\subfloat[][]{\includegraphics[angle=0,width=0.49\textwidth]{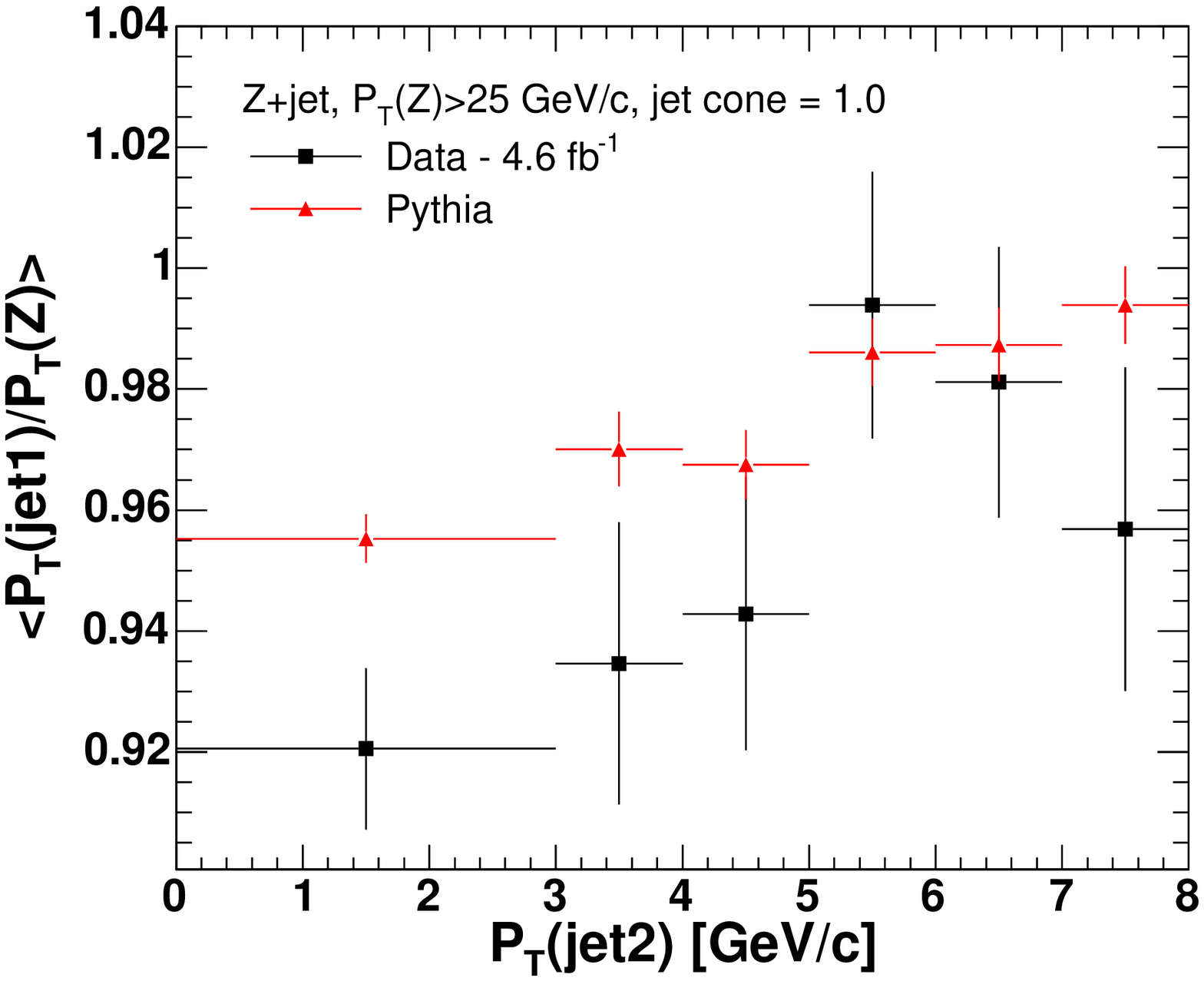}}
\subfloat[][]{\includegraphics[angle=0,width=0.49\textwidth]{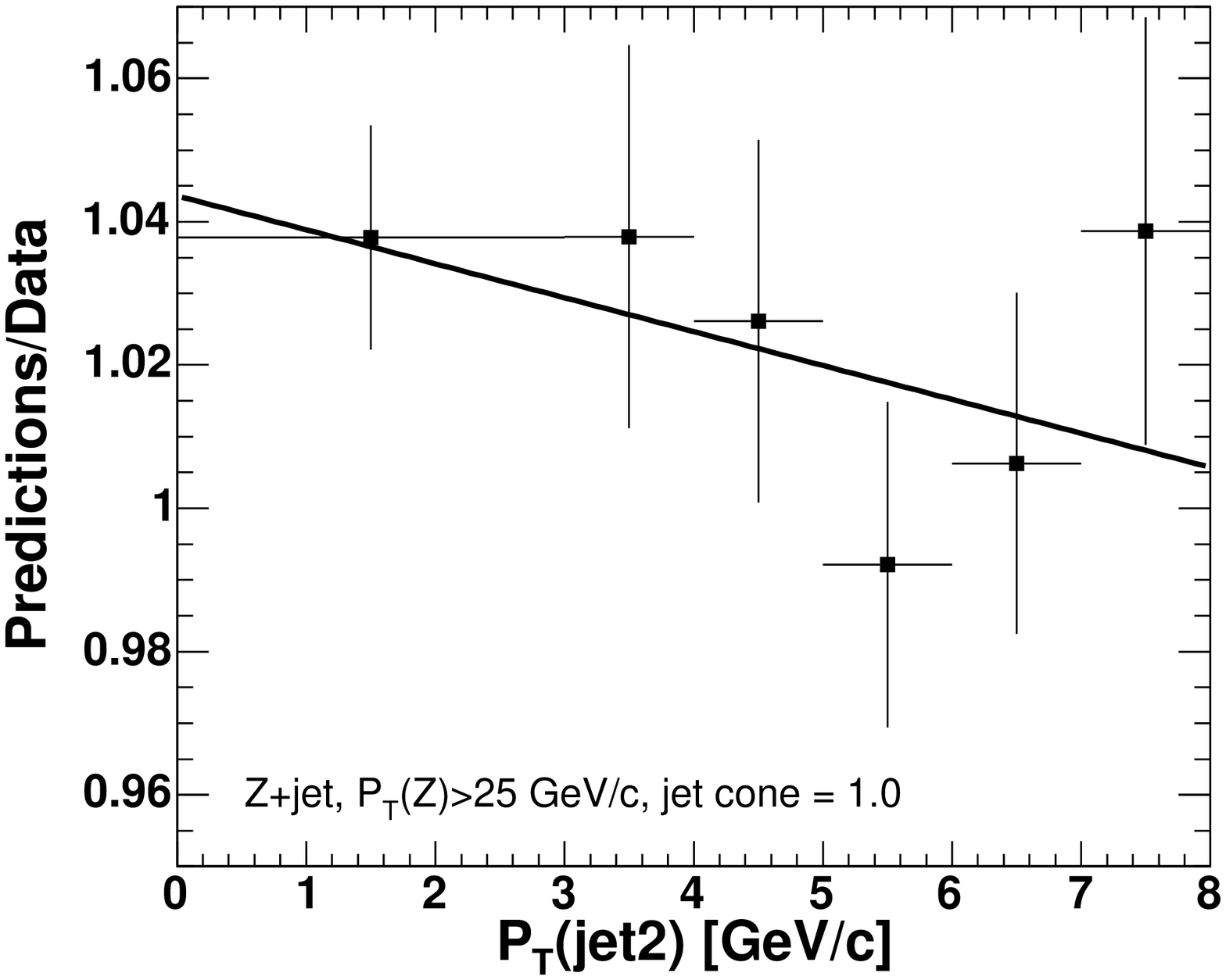}}
\caption{ a.) A comparison of the measured (square markers) and
predicted (triangle markers) \pt-balance as a function of the 2nd jet
\pt~for jets of R=1.0 cone size. The predicted balance is obtained
with {\sc pythia}. The events are required to have only one
interaction per event. b.)  The ratio of predicted to measured
\pt-balance versus the \pt~of the second jet. The linear fit of the
ratio resulted in a slope of -0.47$\pm$0.42 \%/\GeVc.}
\label{fig:j2_pt_cone10}
\end{figure}

 We use the \pt-cutoff of the sub-leading jet to estimate the
variation of the balance due to the out-of-cone radiation. The
agreement between data and predictions improves as we decrease the
cut-off value on \pt$(jet2)$ as shown in Figs.~\ref{fig:j2_pt_cone04},
\ref{fig:j2_pt_cone07}, and \ref{fig:j2_pt_cone10} (all the other
selection requirements are kept the same). The extrapolation to the
point where \pt$(jet2)$ is zero describes the case where both data and
predictions do not have any large-angle FSR.
%

\par We use the ratio of \pt-balance between data and prediction to
estimate the uncertainty due to large-angle FSR. The ratio is obtained
as a function of \pt~of the sub-leading jet, \pt$(jet2)$. To obtain
the systematic uncertainty we multiplied the mean value of \pt$(jet2)$
by the slope of the linear fit of the ratio. The uncertainties are
summarized in Table~\ref{table:systematics}. As an alternative
approach, one can take the difference between the mean value of the
ratio and that obtained using the linear extrapolation to \pt$(jet2)$
$=$ 0~\GeVc as a systematic uncertainty. The latter approach gives
different values for the uncertainty than the first one ( -2.0\%,
+0.4\%, and 2.4\% for the jet cone sizes of 0.4, 0.7, and 1.0,
respectively).

\section{Final state radiation uncertainties}
\label{FSR}
\par Final state radiation (FSR) from unfragmented partons is modeled
by time-like parton showers in {\sc pythia}. The predicted \pt-balance
is sensitive to the rate of FSR. In this study, the variation of FSR
parameters is performed similarly to that for ISR (see
Section~\ref{ISR} and~\cite{top_mass_jes}). To produce the systematics
samples we altered the parameters PARP(72) and PARP(71) in {\sc
pythia}. The value of PARP(72) defines the parameter $\Lambda_{\rm
QCD}$ used in simulation of time-like showers of partons produced in
ISR. By default, the same $\Lambda_{\rm QCD}$ parameter from the PDF
set (CTEQ5L in this case) is used for ISR and FSR partons. The value
of PARP(71) is used as a multiplicative scaling factor to the $Q^2$
scale of the hard scattering to define the maximum parton virtuality
for time-like showers. The variation of \pt-balance in the FSR
systematics samples is about 0.1-0.4\%, as shown in
Table~\ref{table:systematics}.

\section{Multiple proton-proton interactions}
\label{pileup}
\par The \pt-balance is sensitive to multiple \ppbar~interactions
overlapping in-time with the hard process. The number of interactions
per event is estimated by observing additional primary vertices along
the beam-line. The additional interactions are likely to be
minimum-bias collisions.
%

\par The uncertainty in the \pt-balance arises from the present
limited ability of {\sc pythia} to describe the calorimeter energy in
a minimum bias event accurately. The momentum distribution of charged
particles in the predictions, measured in the magnetic spectrometer
(COT), was tuned to data only for particles with \pt $>$
0.5~\GeVc~\cite{Field_UE_Tune} as soft charged particles curl up in
the magnetic field and do not reach the calorimeters. Consequently the
response for soft (\pt $<$ 0.5) neutral hadrons is not measured. The
\pt-balance varies to the number of primary verticies differently in
data and predictions. The difference in variation of the \pt-balance,
on the order of a percent, is taken as a systematic uncertainty (see
Table~\ref{table:systematics}).
\section{Detector simulations for the jet energy response}
\label{det_SPR}
\par The jet energy scale relies on measurements of the response of
the calorimeters to single charged particles whose momenta are
measured precisely using the magnetic spectrometer of CDF as well as
test-beam measurements at the highest momenta~\cite{jet_corr}.
Uncertainties in transferring the measured single-particle response to
a parametric model of the calorimeter jet response contribute
significantly to the uncertainty on the CDF jet energy scale. The net
effect of varying the single particle response on the simulated
\pt-balance is typically 2.5\% as shown in
Table~\ref{table:systematics}.

\section{Summary of systematic uncertainties}
\label{sum_of_syst}
\par We observe a significant discrepancy in the \pt-balance of
$Z$-bosons and single jets between measurements and prediction, with
the data being lower in the \pt~of the jet compared to predictions.
We find that {\sc pythia} predicts the fractions of quark and gluon
jets accurately. In Table~\ref{table:systematics} we summarize the
estimated variations (taken as systematic uncertainties) of the
predicted balance. The totality of the variations is comparable to the
observed discrepancy, with the largest contributions being those from
large-angle FSR and the modeling of the single-particle response of
the calorimeters.
\begin{table}[h]
\centering
\begin{tabular}{|l|c|c|c|}
\hline
Source of uncertainty & R = 0.4 & R = 0.7 & R = 1.0 \\
\hline
renormalization and factorization scales &  +0.9 -0.0 & +0.9 -0.4 &  $\pm$0.4 \\
FSR parameters in {\sc pythia} &  $\pm$0.4 & $\pm$0.1 &  $\pm$0.1 \\
MEs and jet-parton matching &  +0.8 -0.0 & +1.1 -0.0& +0.8 -0.0\\
detector simulations (single particle response) &  $\pm$2.5 & $\pm$2.5 &  $\pm$2.5 \\
multiple proton interactions & +1.0 -0.0 & +1.2 -0.0&  +1.2 -0.0\\
large-angle FSR (limitation of PS) & +0.0 -2.9 & +0.0 -0.2 & +1.7 -0.0\\
\hline
Estimate of the total variation  & +3.0 -3.8 & +3.1 -2.5 &  +3.4  -2.5 \\
\hline
\hline
The observed discrepancy & +4.7 & +3.2 & +2.0 \\
\hline
\end{tabular}
\caption{The effect on the predicted mean \pt-balance of varying
parameters in the modeling and event selection, in percent, for jet
cone sizes R = 0.4, 0.7, and 1.0. The variations are evaluated for
{\sc pythia} events with \pt$(Z)$ $>$ 25~\GeVc. The observed
discrepancy is defined as the \pt-balance in predictions divided by
that in data; the predicted jet energies are higher than those in
data. The discrepancy between data and predictions is comparable with
the estimate of the total variation of the predictions. A positive
variation in the predicted \pt-balance corresponds to an increase in
the jet energies in the MC predictions. The total variation is
calculated by adding the uncertainties in quadrature.}
\label{table:systematics}
\end{table}

\par Data indicates that the large-angle parton radiation is not
modeled accurately by the leading-log parton showering from {\sc
pythia} Tune AW: the MC simulation underestimates the large-angle
radiation. The discrepancy is not addressed by the predictions from
{\sc alpgen + pythia} either; the ME calculations affect only jets
with \pt~above the matching scale (15 \GeVc in the analysis); softer
radiation (\pt(jet2) $<$ 8 \GeVc in the analysis) is simulated via the
parton showering mechanism.

\section{Conclusions}
\label{conclusions}

\par We have estimated the sensitivity of the \pt-balance method to a
number of theoretical factors: the virtuality-ordered parton showering
from {\sc pythia}, tree-level matrix elements, parton distribution
functions, parton-jet matching procedure, renormalization and
factorization scales, multiple \ppbar~interactions, and calorimeter
response of single stable particles. The contribution from each source
of uncertainty is presented in Table~\ref{table:systematics}. The
uncertainty caused by inadequate modeling of the parton shower at
large angles is found to be the largest. The sum of the uncertainties
is consistent with the discrepancy between data and predictions in the
\pt-balance. The remaining uncertainties are significantly
smaller~\cite{jet_corr}.
\par Numerous modern higher-order MC simulations utilize leading-log
parton showering from {\sc
pythia}~\cite{Hamilton:2010wh,Torrielli:2010aw}. The higher-order
calculations of the matrix elements are less sensitive to the choice
of renormalization and factorization scales so that the related
uncertainty should be smaller than that we evaluated. However, the
uncertainty due to large-angle parton radiation is expected to be of
the same magnitude as in the study. We encourage the LHC experiments
to use the distributions in Figs.~\ref{fig:j12_phi_cone04},
\ref{fig:j12_phi_cone07},~\ref{fig:j12_phi_cone10}
and~\ref{fig:j2_pt_cone04},~\ref{fig:j2_pt_cone07},~\ref{fig:j2_pt_cone10}
as a systematic method for tuning the parton showering parameters in
event generators for more accurate jet energy measurements.
%

\begin{acknowledgments}

\par We thank the Fermilab staff and the technical staffs of the
participating institutions for their vital contributions. This work
was supported by the U.S. Department of Energy and National Science
Foundation; the Italian Istituto Nazionale di Fisica Nucleare; the
Ministry of Education, Culture, Sports, Science and Technology of
Japan; the Natural Sciences and Engineering Research Council of
Canada; the National Science Council of the Republic of China; the
Swiss National Science Foundation; the A.P. Sloan Foundation; the
Bundesministerium f\"ur Bildung und Forschung, Germany; the World
Class University Program, the National Research Foundation of Korea;
the Science and Technology Facilities Council and the Royal Society,
UK; the Institut National de Physique Nucleaire et Physique des
Particules/CNRS; the Russian Foundation for Basic Research; the
Ministerio de Ciencia e Innovaci\'{o}n, and Programa
Consolider-Ingenio 2010, Spain; the Slovak R\&D Agency; and the
Academy of Finland. This work has also been supported by the Maria
Goeppert Meyer Fellowship of Argonne National Laboratory, the National
Science Foundation, and the US Department of Energy.

\par We are thankful to David Mietlicki, Ray Culbertson, Peter Skands,
Torbjorn Sjostrand, Alexander Pronko, Larry Nodulman, Willis Sakumoto,
Michelangelo Mangano, Christina Mesropian, Eric J. Feng, and
Jan-Christopher Winter for suggestions.
\end{acknowledgments}

\bibliography{jes_syst_pub}

\begin{thebibliography}{48}
\bibitem [{EtP()}]{EtPt}%
  \BibitemOpen
  \href@noop {} {}\bibinfo {note} {Transverse momentum and energy are defined
  as $\Pt = p\sin\theta$ and $\Et = E\sin\theta$, respectively.}\BibitemShut
  {Stop}%
\bibitem [{CDF(){\natexlab{b}}}]{CDF_coo}%
  \BibitemOpen
  \href@noop {} {} \bibinfo {note} {The CDF coordinate
  system is cylindrical and right-handed, with the $x$ axis horizontal and out
  of the Tevatron ring, the $y$ axis up, the $z$ axis along the proton beam,
  and $r=\sqrt{x^2+y^2}$; $\phi$ is the azimuthal angle. The pseudorapidity
  $\eta\equiv-{\rm ln}(\tan (\theta/2))$, where $\theta$ is the polar angle;
  the $\eta$ regions for detector components are defined with respect to the
  center of the detector. The central region is defined as
  $|\eta|<1$.}\BibitemShut {Stop}%
\bibitem [{ele()}]{electron_track_quality}%
  \BibitemOpen
  \href@noop {} {}\bibinfo {note} {For central electrons at least 5 hits in
  each of 3 axial and 2 stereo layers of the COT are required.}\BibitemShut
  {Stop}%
\bibitem [{had()}]{hadoem}%
  \BibitemOpen
  \href@noop {} {}\bibinfo {note} {The fraction of electromagnetic energy
  allowed to leak into the hadron compartment $\rm E_{had}/E_{em}$ must be less
  than $\rm 0.055\plus0.00045\times E_{em}(\GeV)$ for central electrons, less
  than 0.05 for electrons in the end-plug calorimeters, less than max[0.125,
  $\rm 0.055\plus0.00045\times E_{em}(\GeV)$] for photons.}\BibitemShut {Stop}%
\bibitem [{muo(){\natexlab{a}}}]{muon_track_quality}%
  \BibitemOpen
  \href@noop {} {} \bibinfo {note} {For tight muons at least
  5 hits in each of 3 axial and 3 stereo layers of the COT are required; for
  loose muons with a matching muon stub this is relaxed to 3 axial and 2
  stereo. Loose muons without a matching stub have an additional cut on the
  $\chi ^2$ for the fit to the track.}\BibitemShut {Stop}%
\bibitem [{muo(){\natexlab{b}}}]{muon_cal_cuts}%
  \BibitemOpen
  \href@noop {} {} \bibinfo {note} {The energy deposited in
  the calorimeter tower traversed by the muon must be less than
  $2+\mathrm{max}(0,0.0115\times(p-100))~\GeV $ in the electromagnetic
  compartment and less than $6+\mathrm{max}(0,0.028\times(p-100))~\GeV$ in the
  hadronic compartment.}\BibitemShut {Stop}%
\bibitem [{muo(){\natexlab{c}}}]{muon_stub_matching}%
  \BibitemOpen
  \href@noop {} {} \bibinfo {note} {The muon ``stub'' in the
  muon systems must be within 7, 5, and 6 cm of the extrapolated COT track
  position, in the CMU, CMP, and CMX muon systems, respectively.}\BibitemShut
  {Stop}%
\bibitem [{lep()}]{lepton_scale_factors}%
  \BibitemOpen
  \href@noop {} {}\bibinfo {note} {The maximum correction to the identification
  efficiency for central electrons is 1.7\%, for ``plug'' electrons 6.3\%, and
  for central muons is 7.4\%.}\BibitemShut {Stop}%
\end{thebibliography}

\appendix*
\section{\label{lepton_id}Lepton identification}
\par We use standard CDF definitions for identification (ID) of
electrons and muons as described below~\cite{R_ratio_exp}. The same
lepton ID requirements are applied to events from data and Monte Carlo
simulations.

\par The identification and triggering efficiencies for leptons are
different for events in data and Monte Carlo, although they
demonstrate a very similar energy dependence. To eliminate this
inconsistency we follow the standard CDF practice of using correction
factors (``scale factors'') to re-weight the MC events (see
Section~\ref{MC_lept_corrections}).

\par In order to maintain a high efficiency for $Z$ bosons, for which
we require two identified leptons, we define ``tight'' and ``loose''
selection criteria for both electrons and muons, as described below.

\par To reduce backgrounds from the decays of hadrons produced in
jets, leptons are required to be ``isolated''. The $\Et$ deposited in
the calorimeter towers in a cone in $\eta-\varphi$
space~\cite{CDF_coo} of radius $R=0.4$ around the lepton position is
summed, and the $\Et$ due to the lepton is subtracted. The remaining
$\Et$ is required to be less than 10\% of the lepton $\Et$ for
electrons or $\Pt$ for muons.

\subsection{\label{electron}Electron selection}
\par An electron candidate passing the ``tight'' selection must be
central with $\Et>20$~\GeV, and have: a) a high quality
track~\cite{electron_track_quality} with $\Pt>0.5\cdot\Et$ or $\Pt >
50$~\GeVc; b) a good transverse shower profile at shower maximum that
matches the extrapolated track position; c) a lateral sharing of
energy in the two calorimeter towers containing the electron shower
consistent with that expected; and d) minimal leakage into the hadron
calorimeter~\cite{hadoem}.
\par Additional central electrons, classified as ``loose'' electrons,
are required to satisfy the ``tight'' central electron criteria but
with a track requirement of $\Pt>10$~\GeVc (rather than
$0.5\cdot\Et$), and no requirement on a shower maximum measurement or
lateral energy sharing between calorimeter towers.  Electrons in the
forward calorimeters ($1.2 < |\eta| < 2.5$), also classified as
``loose'' electrons, are required to have $\Et> 20$~\GeV, minimal
leakage into the hadron calorimeter, a track containing at least 3
hits in the silicon tracking system, and a shower transverse shape
consistent with that expected, with a centroid close to the
extrapolated position of the track~\cite{wenu_asymmetry_paper}.

\subsection{Muon selection}
\label{muon}
\par A muon candidate passing the tight cuts must have: a) a well
measured track in the COT~\cite{muon_track_quality} with
$\Pt>20$~\GeVc; b) energy deposited in the calorimeter consistent with
expectations~\cite{muon_cal_cuts}; c) a muon ``stub''~\cite{stub} in
both the CMU and CMP, or in the CMX, consistent with the extrapolated
COT track~\cite{muon_stub_matching}; and d) a COT track fit consistent
with an outgoing particle from a \ppbar~collision and not from an
incoming cosmic ray~\cite{CRAY}.
\par Additional muons, classified as ``loose'', are required to have
$\Pt>20$~\GeVc and to satisfy the same criteria as for tight muons but
with relaxed COT track quality requirements. Alternatively, for muons
outside the muon system fiducial volume, a loose muon must satisfy the
tight muon criteria and an additional more stringent requirement on
track quality, but the requirement that there be a matching ``stub''
in the muon systems is dropped.

\subsection{Corrections due to Modeling of Electrons and Muons}
\label{MC_lept_corrections}
\par Following the standard treatment of lepton efficiencies in CDF,
we re-weight Monte Carlo events to take into account the difference
between the identification efficiencies measured in leptonic $Z$
decays and those used in simulation~\cite{lepton_scale_factors}. We
then make additional corrections for the difference in trigger
efficiencies in simulated events and measured in data. Corrections to
trigger efficiencies are typically 4\% for trigger electrons, 8\% for
trigger muons that traverse both the CMU and CMP systems, and 5\% for
muons in the CMX system. The average weight for \zee~events is 0.939;
for \zmumu~events it is 0.891.
\par We correct the energy of electrons and muons the same way as it
was done for the measurement of the $W$ boson mass~\cite{w_mass}. The
relative positions of the tracker wires were aligned using cosmic
muons. Additional track-level corrections were derived using $W\goes
e\nu$ data to reduce bias between positive and negative particles. The
electron energy was corrected in data for effects due to tower
position and time (aging).

\end{document}